\documentclass[fleqn,11pt,twoside]{article}
\usepackage{amsthm,amsthm,amssymb, color, xcolor,epsfig, graphics, subfigure}
\usepackage{amsmath, graphicx, latexsym, lscape}
\usepackage[utf8]{inputenc}
\usepackage{amsfonts}
\numberwithin{equation}{section}
\usepackage{appendix}
\usepackage{srcltx}
\usepackage{authblk}
\usepackage{float} 
\usepackage{cancel}
\usepackage[left=2.00cm, right=2.00cm, top=2.00cm, bottom=2.00cm]{geometry}
\usepackage{color}
\definecolor{darkgreen}{rgb}{0,0.5,0}
\definecolor{purple}{rgb}{1,0,1}
\newcount\Comments  %
\newcommand{\kibitz}[2]{\ifnum\Comments=1\textcolor{#1}{#2}\fi}

\definecolor{myblue}{rgb}{0,0,0.8}
\definecolor{green}{RGB}{0, 130, 0}
\definecolor{grey}{RGB}{90, 90, 90}

\providecommand*{\dder}[3][]{%
\frac{d^{#1}#2}{d #3^{#1}}}

\makeatletter
\newcommand{\copyrightnote}[2]{{\renewcommand{\thefootnote}{}
 \footnotetext{\small\it
\begin{flushleft}
 \copyright \ #1   #2  
\end{flushleft}}}}

\newcommand{\Name}[1]{\begin{flushleft}
                       \LARGE \bf #1
                       \end{flushleft}\vspace{-3mm}}

\newcommand{\Author}[1]{\begin{flushleft}
                       \it #1 \end{flushleft}}

\newcommand{\Address}[1]{\begin{flushleft}
                       \it #1 \end{flushleft}}

\newcommand{\Date}[1]{\begin{flushleft}
                      \small  \it #1 \end{flushleft}}

\newcommand{\evenhead}{Author \ name}
\newcommand{\oddhead}{Article \ name}

\renewcommand{\@evenhead}{
\hspace*{-3pt}\raisebox{-15pt}[\headheight][0pt]{\vbox{\hbox to \textwidth
{\thepage \hfil \evenhead}\vskip4pt \hrule}}}
\renewcommand{\@oddhead}{
\hspace*{-3pt}\raisebox{-15pt}[\headheight][0pt]{\vbox{\hbox to \textwidth
{\oddhead \hfil \thepage}\vskip4pt\hrule}}}
\renewcommand{\@evenfoot}{}
\renewcommand{\@oddfoot}{}

\long\def\@makecaption#1#2{%
  \vskip\abovecaptionskip
  \sbox\@tempboxa{\small \textbf{#1.}\ \ #2}%
  \ifdim \wd\@tempboxa >\hsize
    {\small \textbf{#1.}\ \ #2}\par
  \else
    \global \@minipagefalse
    \hb@xt@\hsize{\hfil\box\@tempboxa\hfil}%
  \fi
  \vskip\belowcaptionskip}

\newcommand{\JNMPnumberwithin}[3][\arabic]{%
  \@ifundefined{c@#2}{\@nocounterr{#2}}{%
    \@ifundefined{c@#3}{\@nocnterr{#3}}{%
      \@addtoreset{#2}{#3}%
      \@xp\xdef\csname the#2\endcsname{%
        \@xp\@nx\csname the#3\endcsname .\@nx#1{#2}}}}%
}

\newcommand{\resetfootnoterule} {
  \renewcommand\footnoterule{%
  \kern-3\p@
  \hrule\@width.4\columnwidth
  \kern2.6\p@}
}

\renewcommand{\footnoterule}{}

\makeatother

\theoremstyle{definition}

\begin{document}

\renewcommand{\evenhead}{ {\LARGE\textcolor{blue!10!black!40!green}{{\sf \ \ \ ]ocnmp[}}}\strut\hfill 
Y F Adans, H Aratyn, J F Gomes  and G V Lobo
}
\renewcommand{\oddhead}{ {\LARGE\textcolor{blue!10!black!40!green}{{\sf ]ocnmp[}}}\ \ \ \ \  
Volterra-Bogoyavlensky lattices and solutions of Painlev\'e  equations
}

\thispagestyle{empty}
\newcommand{\FistPageHead}[3]{
\begin{flushleft}
\raisebox{8mm}[0pt][0pt]
{\footnotesize \sf
\parbox{150mm}{{\textcolor{blue!10!black!40!green}{{\bf Open Communications in Nonlinear Mathematical Physics}}}
\ \ {Special Issue: Hietarinta}, 2026\\[0.1cm]
\strut\hfill 
ocnmp:18549
pp #2\hfill {\sc #3}}}\vspace{-13mm}
\end{flushleft}}

\FistPageHead{1}{\pageref{firstpage}--\pageref{lastpage}}{ \ \ }

\strut\hfill

\strut\hfill

\copyrightnote{The authors. Distributed under a Creative Commons Attribution 4.0 International License}

\begin{center}

{\bf {\large A Special OCNMP Issue in Honour of Jarmo Hietarinta}}\\[0.2cm]
{\bf {\large on the Occasion of his 80th Birthday}}
\end{center}

\smallskip

\Name{Volterra-Bogoyavlensky lattices
and solutions of  $A_{2n}^{(1)}$ invariant
Painlev\'e 
equations}

\author[1]{J.F. Gomes} 
\author[1]{G.V. Lobo}

\Author{Y. F. Adans, H. Aratyn, J.F. Gomes and G.V. Lobo}

\Address{Y. F. Adans, J.F. Gomes and G.V. Lobo: Universidade Estadual Paulista (Unesp), 
Instituto de F\'{i}sica Te\'{o}rica (IFT), S\~{a}o Paulo, 
Rua Dr. Bento Teobaldo Ferraz 271, 01140-070, S\~{a}o Paulo, SP, Brasil
 }
\Address{H. Aratyn: Department of Physics, 
	University of Illinois Chicago, 
	845 W. Taylor St.
	Chicago, Illinois 60607-7059, USA}

\Date{Received June 18, 2026; Accepted September 22, 2026}

\setcounter{equation}{0}

\smallskip

\noindent
{\bf Citation format for this Article:}\newline
Y F Adans, H Aratyn, J F Gomes and G V Lobo,
Volterra-Bogoyavlensky lattices and solutions of $A^{(1)}_{2n}$
 invariant Painlev\'e  equations
{\it Open Commun. Nonlinear Math. Phys.}, Special Issue:\,Hietarinta, ocnmp:18549, \pageref{firstpage}--\pageref{lastpage}, 2026.

\strut\hfill

\noindent
{\bf The permanent Digital Object Identifier (DOI) for this Article:}\newline
{\it 10.46298/ocnmp.18549}
\strut\hfill


\begin{abstract}
\noindent 
The objective of this work is to develop a framework that exploits 
the lattice structure of the $k$-th
Volterra--Bogoyavlensky equations ($k\in\mathbb N$, $k>1$) to generate
rational solutions of higher symmetric Painlev\'e equations.

For $k=2$, we show that the Volterra lattice, equipped with suitable
initial conditions, exactly models the one- and two-dimensional orbits
generated by half-translation operators of the $A_2^{(1)}$ symmetric
Painlev\'e IV equations. This correspondence yields explicit closed-form
expressions for all solution components in terms of generalized
Okamoto polynomials and leads to new algebraic recurrence relations
among these polynomials.

We present  two generalizations of the above Volterra lattice. One 
is derived from a fractional translation of  the $A_{4}^{(1)}$ 
symmetric Painlev\'e equations. It generalizes Volterra lattice structure in
the multi-component setup of the affine $A_{4}^{(1)}$  group and it is
shown to generate solutions of the $A_{4}^{(1)}$ 
symmetric Painlev\'e equations from the seed solution invariant under
dihedral group $D_{5}$. The other is the $k=3$ Bogoyavlensky
lattice structure. It satisfies recurrence relations that naturally
extend recurrence relations of the Volterra lattice.

These results shed light on connection between
Volterra--Bogoyavlensky lattices, dihedral symmetries, and rational
solutions of higher Painlev\'e  systems.
\end{abstract}

\label{firstpage}

\section{Introduction}
\label{section:introduction}

The principal new result of this paper is the identification of
Volterra lattice equations with orbits of class of
rational solutions of symmetric Painlev\'e
systems. This correspondence
involves orbits generated by
fractional roots of translation operators of symmetric Painlev\'e
systems. The construction provides a systematic
derivation of rational
solutions from dihedral symmetric seed solutions and leads naturally
to  novel algebraic
recurrence relations for generalized Okamoto polynomials.
There are several general concepts that play  key roles 
in the proposed formalism. One has to do with 
dihedral symmetry that provides the organizing principle behind 
the entire construction.

For the $A_n^{(1)}$ symmetric Painlev\'e  equations it is helpful
to analyze their structure  by emphasizing invariance under
dihedral group $D_{n+1}$ of automorphisms. 

Formally, the symmetry group of the $A^{(1)}_n$ invariant Painlev\'e equations
can be viewed as a semi-direct product of the affine Weyl
group  $W(A^{(1)}_n)$, generated by the B\"acklund transformations
$s_i$,  and the dihedral group $D_{n}$ acting by automorphisms
{This is a picture which first emerged in \cite{okamoto} for the
case of Painlev\'e IV based on 
discussion of full automorphism group associated  to the root lattice 
for $A^{(1)}_2$ and was also derived from geometrical approach in
Adler's papers \cite{Adler-polygon,Adlerchain} and in the bilinear
approach to Painlev\'e IV formalism in reference \cite{hietarinta1} }

 \begin{figure}[H]
   \centering
 \scalebox{2.7}{ \includegraphics[width=5cm]{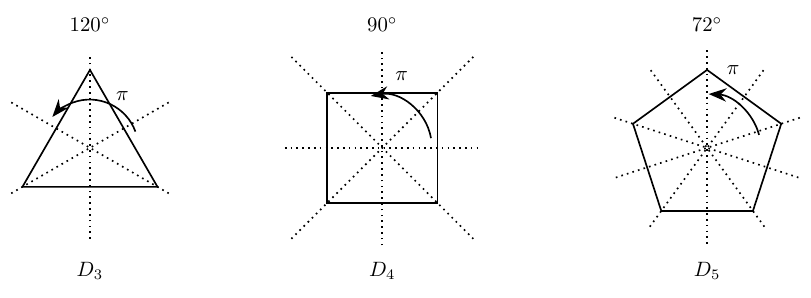}}
   \caption{$n$-polygons for $n=3,4,5$ with their axes of reflections
   and basic rotations by $360^{0}/n$. They correspond to group of automorphisms 
   of ${A}_{2}^{(1)}, { A}_{3}^{(1)}, { A}_{4}^{(1)}$
   invariant Painlev\'e equations. 
For $n=3$ we deal with the  symmetric Painlev\'e  IV equations
and dihedral symmetry of a triangle.
}
       \label{fig:polygons}
 \end{figure}

This is illustrated by figure \ref{fig:polygons} that 
shows  closed polygons with  $3,4$ and $5$
vertices with their symmetry axes and basic rotations for
${A}_{2}^{(1)}, {A}_{3}^{(1)}, {A}_{4}^{(1)}$ invariant
Painlev\'e equations. 
By placing B\"acklund transformations $ s_i$ in the corresponding
polygon's vertices labeled by $i$ we can schematically find out how  
$s_i$ will
transform under reflections and rotations of $D_k$ acting as
a group of automorphisms.
The  $4$-th closed polygon is a square and we have shown in
\cite{p5degegneracy}  that
the dihedral symmetry $D_4$ in such case contains reflections that have 
fixed points. Presence of these fixed points is 
responsible for the two-fold degeneracy of $A_3^{(1)}$
invariant Painlev\'e equations.
The solutions of the corresponding 
symmetric Painlev\'e V  equations were constructed in \cite{p5degegneracy} 
by actions of two abelian translation operators from symmetric
seed solutions. 

We introduce the $k$-th Volterra-Bogoyavlensky
lattice equations in section \ref{section:VB} and derive the 
bilinear recurrence relations that under proper initial conditions
are equivalent to the original Volterra-Bogoyavlensky lattice equations
but allow for construction of  recursively higher polynomials from
which we eventually build the rational solutions.

For the dihedral groups with odd number of symmetry axes  there are no fixed points.
For example for $k=2$ we deal with the  symmetric Painlev\'e  IV equations
and a dihedral symmetry of a triangle. 
In section \ref{section:P4} a dihedral symmetry of a triangle 
is analyzed in terms of square roots
of translations that no longer  are abelian. 

In section \ref{section:double}, acting with two different 
half-translations belonging to two different directions within the
$A_2^{(1)}$ group leads to Volterra orbits enumerated by two integers
and naturally introduces generalized Okamoto polynomials that
give a complete and novel description for all three components of $f_1,f_2,f_3$
solutions in a closed form.
We also uncover interesting relations between generalized Okamoto
polynomials that including some new algebraic recurrence relations.

Continuing in section \ref{section:double} we provide formulation  of Volterra lattice 
and double Volterra lattice in which they exactly model the orbits 
generated by actions of square-roots translation operators. 
The (single) Volterra
lattice  models the action
by one of the  square-roots of translation operators acting on the
initial value of $x$. 
The double Volterra lattice models the action
by two  (necessarily  neighboring) square-roots of translation operators
$T_i^{(m/2)} T_{i+1}^{(n/2)}, n,m \in \mathbb{N}$.
Accordingly,
 these solutions are parameterized by two positive integers
 that enter the formalism as powers of the square-roots of 
 translation operators providing explanation for the structure behind
 the generalized Okamoto polynomials.
We will show that it is possible to arrange points on a 
Volterra lattice so they correspond
to various gauge variants of triplet being a solution to the  symmetric 
Painlev\'e  IV equations when acted on by half-translations 
(see \cite{AFGZ} for the origin of such connection within the AKNS model).

The negative Volterra lattice is introduced in section \ref{section:negative}
and understood as generated from the seed solutions $f_i=-x$ that also are
dihedral symmetric.

In section \ref{section:k3Bogo}, 
we  realize Volterra structure from the fourth root
of translation operator
 within the framework of $A_{4}^{(1)}$ symmetric Painlev\'e equations, {which first appeared in literature  in \cite{Adler-polygon,Adlerchain,noumi,noumibk}}.

We also describe the orbit 
of the $k=3$ Bogoyavlensky lattice and show how to solve the
underlying recurrence relations for special initial conditions.

The discussion of results and an outlook is given in section
\ref{section:discussion}.

Finally, the two appendices collect some concrete expressions for Okamoto 
polynomials and their determinant representations.

\section{Volterra-Bogoyavlensky lattices and bilinear recurrence
relations}
\label{section:VB}
We define the 
$k$-th Volterra-Bogoyavlensky lattice equations \cite{Bogo} as 
\begin{equation}
\dder{F_n}{x}=F_n  \sum_{i=1}^{k-1}\left( F_{n+i} -  F_{n-i} \right)
, \quad k=2, 3, {\ldots} 
\label{Bogoeqs}
\end{equation}
The $k$-th Bogoyavlensky lattice with
appropriately chosen boundary conditions generalizes the $k=2$ Volterra lattice
build of the $k$-th order root of translation operators on 
$A_{2 (k-1)}^{(1)}$ Painlev\'e equations. 
This symmetry is such 
that it generates rational solutions out
of the seed solution uniquely selected to be invariant  under a dihedral
symmetry of the $k+1$-polygon. 

For both Volterra and the $k$-th Bogoyavlensky lattice
we choose initial  conditions to coincide with the
seed solutions determined by
the following configuration 
$f_i^{(0)}=x, \alpha_i=1, i=1,2, {\ldots} , k+1$. 

For $k=2$ the equation \eqref{Bogoeqs} becomes the well-known Volterra
equation, 
\begin{equation} \mu_n^{\prime}= {\mu_n}(\mu_{n+1}-\mu_{n-1}), \;\; \;\;n=0,1,2,{\ldots} ,
\label{unVolterraeqs} \end{equation}
which will be shown to coincide fully with the action of 
half-translation  on $A_2^{(1)}$ symmetric  Painlev\'e equations (also 
known as symmetric Painlev\'e IV equations) that  generate rational
solutions from the dihedral symmetric seed solutions.

For $k=3$ we have  equation
\begin{equation}
\dder{F_n}{x}=F_n  \left( F_{n+1}+F_{n+2}  -  F_{n-1}-F_{n-2} \right)
\label{Bogok3}
\end{equation}
and we will explore its connection to orbits on the 
$A_4^{(1)}$ symmetric Painlev\'e equations as will be shown below.

We show that the generalized Volterra lattice 
obtained from fourth order root of the translation operator 
 of the $A_4^{(1)}$ symmetric Painlev\'e equations 
will generate rational solutions of these equations.

We will now introduce a sequence of functions $\Omega_n$ defined through relation:
\begin{equation}
F_n=\Bigg( \ln \frac{\Omega_n}{\Omega_{n-1}} \Bigg)^{\prime}\,.
\label{OmegaF}
\end{equation}
Inserting this expression on the right hand side of
Bogoyavlensky equation \eqref{Bogoeqs}
we obtain after a simple algebra:
\[
\sum_{i=1}^{k-1}\left( F_{n+i} -  F_{n-i} \right)=
\left(\ln
\frac{\Omega_{n+k} \, 
\Omega_{n-(k-1)}}{\Omega_n  \,\Omega_{n-1}}\right)^{\prime}\,.\]
For $F_n$ that satisfies equation \eqref{Bogoeqs} the above expression
is equal to ${F_n^{\prime}}/{F_n}$. After 
integration  and ignoring the integration constants, we
arrive  at an
alternative representation for $F_n$:
\begin{equation}
F_n  =\frac{\Omega_{n+k}\Omega_{n-(k-1)}}{\Omega_n  \,\Omega_{n-1}}
\,.
\label{Bundef2}
\end{equation}
Comparing the two expressions for $F_n$ given in \eqref{OmegaF}
and \eqref{Bundef2} and shifting $n \to n+1$ we arrive at a basic
recurrence for $\Omega$'s:
\begin{equation}
 \Omega_{n+k}\Omega_{n-(k-1)}=
 \Omega_{n+1}^{\prime}\Omega_n- 
\Omega_n^{\prime} \Omega_{n+1}\, . 
\label{BCBogo}
\end{equation}
This expression simplifies to  recursive relations for  $k=2$ and $k=3$
lattices:
\begin{align}
\Omega_{n-1} \Omega_{n+2}&=\Omega_n \Omega_{n+1}^{\prime}- 
\Omega_n^{\prime} \Omega_{n+1}\, , 
 \;\; n=0,1,2,{\ldots} \,,
\label{BCuniOmega}\\
\Omega_{n+3}\Omega_{n-2}&= \Omega_n \Omega_{n+1}^{\prime}- 
\Omega_n^{\prime} \Omega_{n+1}\, ,  \;\; n=0,1,2,{\ldots} \,,
\label{BCBogo3}
\end{align}
We will use these recurrence relations to obtain the associated polynomials involved
in $\Omega_n$'s for $k=2$ and $k=3$ Volterra-Bogoyavlensky lattices.

\section{Symmetric Painlev\'e IV Equations and invariance under $D_3$}
\label{section:P4}
We  work with the  $A_2^{(1)}$ invariant  Painlev\'e IV equations $g_i'=g_i
(g_{i+1}-g_{i-1})+\beta_i, i
\in  \mathbb{Z}/3 \mathbb{Z}$, (integers mod $3$).  Explicitly they
read  as
\begin{equation}
g_1'=g_1 \left(g_2-g_3\right)+\beta_1\, , \;\;
g_2'=g_2\left(g_3-g_1\right)+\beta_2\, , \;\;
g_3'=g_3 \left(g_1-g_2\right)+\beta_3\, ,
\label{symp4}
\end{equation}
where $g_i = g_i(z)$ and $'=d/dz$. 
We impose conditions :
\begin{equation}
g_1+g_2+g_3 =\sigma z\, , \quad \beta_1+\beta_2+\beta_3=\sigma {.}
\label{alpha0f0}
\end{equation}
on $g_i, \beta_i, i=1,2,3$ that are consistent with the relation $\sum_i g_i'=
\sum_i \beta_i$ that follows from equation \eqref{symp4}.
Here $\sigma$ is an additional parameter. {Such additional parameter 
is not uncommon in various 
presentations of Painlev\'e IV  equation, see e.g.
\cite{noumi,hietarinta1,hietarinta}.
In this work adding a $\sigma$ parameter  enhances a parameter space to
 enable a manifest and transparent inclusion 
of the dihedral symmetry group $D_3$ \cite{AAGZ}. Especially
reflections of $D_3$ are simply realized as exchange of two
 vertices of a triangle  around the reflection axis.}
In this  setup we  consider the seed solution: 
\begin{equation}
 g_i= \frac{\sigma\, z}{3}, \quad \beta_i= \frac{\sigma}{3} \, ,
\label{symsol}
\end{equation} to equations \eqref{symp4} and to
the normalization condition \eqref{alpha0f0} that is 
manifestly symmetric  under  $D_3$ reflections and
rotations. 

Equations \eqref{symp4} are manifestly invariant under B\"acklund transformations
$s_i$ ($i=1,2,3$) and automorphism $\pi$ that  form the extended affine Weyl 
group $A_2^{(1)}$ \cite{noumi}. These transformations satisfy 
relations.
\begin{equation}
s_i^2=1,\quad (s_is_{i+1})^3=1,\quad 
\end{equation}

{
Following \cite{AAGZ} we next employ a new 
parameter $\sigma$ for  
a simple construction of additional automorphisms  
$ \pi_i, i=1,2,3$  :
\begin{equation}
\begin{array}{c|ccc|ccc|c|c|}
{} & {\beta_3} & { \beta_1} &
{ \beta_2} & { g_3} & { g_1} &
{ g_2}& \sigma & z \\
\hline
{ \pi_3}&{ -\beta_3} &{
 -\beta_2} &{ -\beta_1} &{
 g_3} &{ g_2} &{ g_1}&-\sigma &-z \\
\hline
{ \pi_1}&{ -\beta_2} &{
 -\beta_1} &{ -\beta_3} &{
 g_2} &{ g_1} &{ g_3}&-\sigma &-z \\
\hline
{ \pi_2}&{ -\beta_1} &{
 -\beta_3} &{ -\beta_2} &{
 g_1} &{ g_3} &{ g_2}&-\sigma &-z  
\end{array}
\label{rhoiss}
\end{equation}
that act on $f_i,\alpha_i, i=1,2,3$ through a simple exchange of
the triangle vertices $i-1,i+1$ while keeping
symmetric P$_{\rm IV}$ equations  \eqref{symp4} invariant.
For that reason the additional automorphisms 
$ \pi_i, \,
i=1,2,3 $ can be identified with reflections and
together with the cyclic rotation $\pi$ 
form the $D_3$ dihedral group.
}

{
In references \cite{Adler-polygon,Adlerchain}
the dihedral group $D_3$ was explicitly constructed from geometrical
considerations, while in reference \cite{hietarinta1} it was obtained 
within the setting of bilinear approach to the
symmetric Painlev\'e IV equations. In all these references 
 it was shown that the group elements
of $D_3$ 
agree with 
automorphisms
of the Dynkin diagram of $A^{(1)}_2$. Their construction 
was presented using the variable which we will introduce below in
equation \eqref{fromztox}.}

\begin{figure}[H]
   \centering
 \scalebox{1.7}{ \includegraphics[width=4cm]{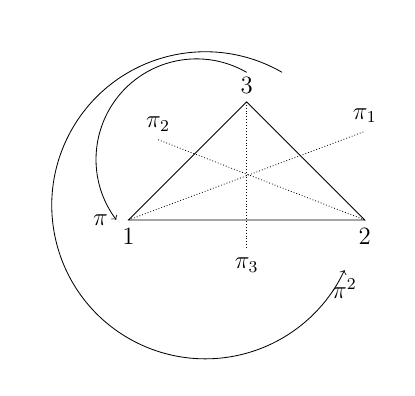}}
   \caption{Reflections $\pi_1,\pi_2,\pi_3$ and rotations
   $\pi, \pi^2$ of $D_3$. Insert $s_i,f_i$ at a vertex $i$ to see
   how it transforms.}  
     \label{fig:triangle2}
 \end{figure}
$D_3$ is a special case of 
$D_{2n+1}$ dihedral group consisting of 
dihedral symmetries preserving the $(2n+1)$-polygon for $n=1,2,3,{\ldots} $.

All automorphisms $\pi_i$ square to one:
\begin{equation}
 \pi_i^2=1, \quad i=1,2,3 \, ,
\label{pisquare}
\end{equation}
and satisfy 
the so-called braid relations 
\begin{equation}
\pi_i \pi_j \pi_i =\pi_j \pi_i \pi_j, 
\quad i \ne j \, {.}
\label{braidrelations}
\end{equation}
Furthermore,  their composition reproduces the rotations:
\begin{equation}
\pi= \pi_i \pi_{i+1} ,\;\;\;\; \pi^{-1}= \pi^2= \pi_{i+1}
\pi_i,\;\;\;\; i=1,2,3 \, .
\label{braid2}
\end{equation}
The group of affine B\"acklund transformations can be written 
as 
\[\begin{split} {\widetilde W}(A_2^{(1)}) &= D_3 \ltimes  W(A_2^{(1)})\\
W(A_2^{(1)})&= \langle s_1,s_2,s_3 \rangle, \quad
s_i^2,\, i=1,2,3\;\; s_i s_j s_i=s_js_i s_j , \; i = j \pm 1
\end{split}
\]
The dihedral group acting transformations $s_i$ via the following
automorphisms
\[\pi^3=1,\quad \pi
s_i=s_{i+1}\pi,\quad 
\pi_i s_i=s_i \pi_i,\; \pi_i s_{i+1}=s_{i-1} \pi_i.
\;\;i =1,2,3 \, .
\]
Generalizing this set-up to the $A^{(1)}_{2n}$ Painlev\'e equations
one encounters a symmetry group 
consisting of the semi-direct product of the Weyl symmetry group of  
$s_1,{\ldots} ,s_{2n+1}$ transformations with the dihedral group $D_{2n+1}$.

For more facts on  $A_{2}^{(1)}$ and $A_{4}^{(1)}$ 
invariant  Painlev\'e equations and
their symmetries  see e.g. \cite{noumi, noumibk}. Origin of many symmetric
Painlev\'e equations from
a periodic sequence of Darboux transformations for a 
Schr\"odinger problem is explained in \cite{hietarinta}.


There are three translation operators associated 
with $A^{(1)}_{2}$, which are given by:
\begin{equation}
T_1=\pi s_2 s_1,\quad T_2= s_1\pi s_2, \quad
T_3= s_2 s_1\pi \, .
\label{Tidefs}
\end{equation}
The translation operators commute among themselves. They are not 
independent as they satisfy the product relation
$T_1 T_2 T_3=1$.

It turns out that for $A_{2}^{(1)}$
the symmetry structure is fully revealed by
considering the square-roots of the abelian translations operators
\begin{equation}
T_1^{1/2}= \pi^{-1} s_1,\quad T_2^{1/2}= s_1\pi^{-1}, \quad
T_3^{1/2}= s_2 \pi^{-1} \, ,
\label{squaresTi}
\end{equation}
We will refer to them as half translations. Their squares  reproduce 
translations but they
no longer commute, instead they satisfy the braid relations \cite{AGLZ-braids}:
\[
T^{1/2}_i T^{1/2}_jT^{1/2}_i=T^{1/2}_j T^{1/2}_i T^{1/2}_j \, .
\]
In this project solutions to symmetric Painlev\'e IV equations are 
obtained by acting with square-roots of translation operators on the
fully
dihedral symmetric seed solution. We will show that their actions
form the Volterra lattice.

\subsection{Change of variables $z\to x$}
To simplify the comparison with the Volterra lattice construction and
streamline the discussion we will
change the underlying variable $z \to x$ and make a corresponding
redefinition of $g_i$. In terms of new variables
the relations satisfied by the seed solution will simplify and 
read :
\begin{equation}
{ f}_i =x, \alpha_i=1,  \quad \sum_{i=1}^3 { f}_i = 3 x ,\;\;\, \sum_{i=1}^3 
\alpha_i=3, , 
\label{seedx}
\end{equation}
Another dihedral symmetric solution is
\begin{equation}
{ f}_i =-x, \,\; \alpha_i=-1 
\label{seedmx}
\end{equation}
which will give rise to the negative Volterra lattice.

Let us assume for the moment 
that $\sigma $ is positive and introduce the
following change of variables
\begin{equation}
x = z \, \sqrt{\frac{\sigma }{3}} , \quad { f}_i=g_i\,
\sqrt{\frac{3}{\sigma }} \,.
\label{fromztox}
\end{equation}
It follows that ${ f}_i $ will satisfy \eqref{seedx} for $g_i$
that satisfies \eqref{symsol}.
Correspondingly 
we find
\[
\frac{\sigma }{3} \frac{d} { d x}  { f}_i=
\frac{\sigma }{3}  { f}_i ( { f}_{i+1}- {
f}_{i-1})+\beta_i\, ,
\]
or  using that $d / dz= \sqrt{\sigma/3}\; d / dx$ and substituting
${ \alpha}_i =
  \frac{3}{\sigma }\beta_i$ :
\begin{equation}
\frac{d} { d x}  { f}_i=
  { f}_i ( { f}_{i+1}- { f}_{i-1})+
  { \alpha}_i ,\;\; i=1,2,3  \, ,
\label{xsymmP4}
\end{equation}
We note that $\sum_i { \alpha}_i =3$,  which is  consistent with
relation \eqref{seedx} and also with $\sum_i \beta_i=\sigma$.

Each dihedral reflection $\pi_i$ transforms $z \to -z, \sigma \to -\sigma$
that results in $x \to - x\sqrt{-1}$ and ${ f}_i (x) \to  {
f}_i(i x) /\sqrt{-1} $ and ${ f}_{i\pm 1}(x) \to { f}_{i\mp 1}(ix)/\sqrt{-1}$ .
Also ${\alpha}_i \to  {
\alpha}_i $ and ${ \alpha}_{i\pm 1} \to { \alpha}_{i\mp 1}$.
Writing $\sqrt{-1}=\pm i$ we have:
\begin{equation}
\begin{split}
\pi_i &: x\to {\bar x}= - (\pm) ix =\mp i  x,\;\;
{ f}_i (x)  \to \mp i { f}_i ({\bar x}), \;
{ f}_{i\pm 1}(x) \to \mp i { f}_{i\mp 1}({\bar x})\\
\pi_i &: { \alpha_i} \to { \alpha_i}, \;\;
{\alpha}_{i\pm 1} \to {\alpha}_{i\mp 1} 
\label{piifx}
\end{split}
\end{equation}
{We recognize in these transformations 
the dihedral reflections given 
in references \cite{Adler-polygon, Adlerchain,hietarinta1}. These symmetries 
keep equations \eqref{xsymmP4} invariant 
while they transform the argument $x$ to $\mp i x$ (or $x^2$ to
$-x^2$). See also \cite{folding}, where such symmetries appeared in
the setting of the  Painlev\'e IV Hamilton equation.}
Note that there is
an ambiguity in sign (an  immaterial  sign in front of ``$i$'') that is 
present in the
formula. These transformations will be shown to
transform the underlying Volterra lattice to the ``negative'' Volterra
lattice.




\subsection{Half-translations acting on equations \eqref{xsymmP4}}

In the setting of equations \eqref{xsymmP4}
the square root translation operators \eqref{squaresTi}
act as follows:
\begin{xalignat}{3}    
T^{\frac12 }_3 (f_3)&= f_2+\left(\ln f_3\right)_x, & 
T^{\frac12 }_2 (f_1)&= f_2, &
T^{\frac12 }_1( f_2)& = f_2 -\left(\ln f_1\right)_x\nonumber\\
T^{\frac12 }_3 (f_1)&= f_1 -\left(\ln f_3\right)_x, & T^{\frac12 }_2 (f_2)&= f_1 +\left(\ln f_2\right)_x, &
T^{\frac12 }_1( f_3)& = f_1 \label{gdef1}\\
T^{\frac12 }_3 (f_2)&= f_3, & T^{\frac12 }_2 (f_3)&= f_3-\left(\ln f_2\right)_x, &
T^{\frac12 }_1( f_1)& = f_3 +\left(\ln f_1\right)_x \, .\nonumber
\end{xalignat}
We adopt the following notation:
was in\begin{equation} 
{f}_{i,\, 0} = (x,x,x),\;\;  { \alpha}_{i,\, 0}= (1,1,1)\, .
\label{inidef}
\end{equation}
Then for repeating $p$-times the $T_1^{\frac12}$ transformations 
leads to :
\begin{equation} \begin{split}
\left({ f}_{1, \, p} , { f}_{2,\,p} , { f}_{3,\,p} \right)
&= \left( (T_1^{\frac12})^p ({f}_{1,\,0}) ,  ( T_1^{\frac12})^p
({f}_{2,\,0}), (T_1^{\frac12})^p ({f}_{3,\,0} ) \right)
\\&=\left( T_1^{\frac{p}{2}}({ f}_{1,\,0}) , T_1^{\frac{p}{2}}({
f}_{2,\,0} ) ,T_1^{\frac{p}{2}}({ f}_{3,\,0}) \right)\, .
\label{fip}
\end{split}\end{equation}
From equation  \eqref{gdef1} we deduce that it holds that 
${ f}_{3, \, n}= { f}_{1, \,n-1}$.

For odd indices $p=2n+1$  the functions ${ f}_{i, \,2n+1}= 
(T_1^{\frac12 })^{2n+1}
({ f}_{i,\, 0})$ satisfy the symmetric
Painlev\'e IV equation with the parameters
\begin{equation}
{ \alpha}_{i, \,2n+1}= (2 +3n , 2,
-1-3n) , \quad n=0,1,2,3,{\ldots} 
\label{alpha1sodd}
\end{equation}
and transform $2n \to 2n+1$ as
\begin{equation}\begin{split}
{ f}_{1, \,2n+1}&={ f}_{2,\, 2n}+ \frac{3n+1}{ f_{1,\,2n} },
\qquad n=0, 1,2,{\ldots} \\ { f}_{2, \, 2n+1}&={ f}_{3,\, 2n}- 
\frac{3n+1}{ f_{1, \, 2n} },\quad
{ f}_{3, \, 2n+1} ={ f}_{1, \, 2n}\, .
\end{split}
\label{T14kpso}
\end{equation}
For  even upper indices  the $T_1^{\frac12 }$ transformation $2n-1 \to
2n$ is
\begin{equation}\begin{split}
{ f}_{1, \,2n}&={ f}_{2, \, 2n-1}+ \frac{3n-1}{ f_{1,\,2n-1}}
,\qquad n=0,1,2,{\ldots} \\
{ f}_{2, \, 2n}&={ f}_{3, \, 2n-1}-\frac{3n-1}{{
f}_{1,\, 2n-1}},\quad
{ f}_{3, \,2n}={ f}_{1,\, 2n-1}
\end{split}\label{T14kse}
\end{equation}
and functions ${ f}_{i, \,2n}$ satisfy the symmetric
Painlev\'e equations with the parameters
\begin{equation}
{\alpha}_{i, \,2n}= (1 +3n, 1,1 -3n)\, .
\label{alpha1seven}
\end{equation}

 We can rewrite action of $T_1^{\frac12 }$ given in 
relation \eqref{gdef1} as the first order differential
recurrence relations
\begin{align} 
{ f}_{1, \, n}&= { f}_{1, \, n-2}+ \left( { f}_{1,\,n-1} \right)^{-1}
\frac{d}{d x}  { f}_{1,\, n-1} \label{mdiffg2f1}\\
{ f}_{2,\, n}&= { f}_{2,\, n-1}-\left( { f}_{1, \, n-1} \right)^{-1}
\frac{d}{d x}  { f}_{1, \, n-1} \, .\label{mdiffg2f2}
\end{align}
Adapting the notation $\mu_n=f_{1, \, n}, \nu_n=f_{2, \,n}$ we 
recognize in relation \eqref{mdiffg2f1} Volterra equation
\eqref{unVolterraeqs} for $\mu$
It  also follows  that $
\mu_0=x$ and  $\mu_{-1}=x$ since ${ f}_{1, \,n= -1}={
f}_{3, \,n= 0}=x$. Thus the Volterra lattice equations
\eqref{unVolterraeqs} with the initial conditions 
$\mu_0=x=\mu_{-1}$, models the action of half translations of the
symmetric Painlev\'e IV equations.


Using equations \eqref{alpha1sodd}-\eqref{alpha1seven}
we can rewrite Volterra orbit
of Painlev\'e IV equations as recurrence relations:
\begin{align}
\mu_{n+1}&=\frac12 \left( \frac{\mu_{n}'}{ \mu_{n}}+3 x -
\mu_{n}+ \frac{C_n}{ \mu_{n}} \right), \;\; n=0,1,2,{\ldots} 
\label{f1recuru}\\
\nu_{n} &=  \mu_{n+1}- \frac{C_n}{ \mu_{n}} , \;\; n=0,1,2,{\ldots}
 \nonumber\\
&=  \mu_{n-2}- \frac{C_{n-1}}{ \mu_{n-1}} ,  \;\; n=0,1,2,{\ldots}
\label{f2recuru}
\end{align}
Since $\nu_{n} +\mu_n+\mu_{n-1}=3x$ we obtain a recurrence relation
\begin{equation}
\mu_{n+1}= 3 x+\frac{C_n}{ \mu_{n}} -\mu_n-\mu_{n-1} \, .
\label{2recu}
\end{equation}
The initial conditions for recursion relations 
\eqref{f1recuru}, \eqref{f2recuru}
are :
\begin{equation}
\mu_0=x\;\;\quad \nu_0=x\,.
\label{inicmunu}
\end{equation}
The condition $\nu_0=x$ is consistent with the first of relations
\eqref{f2recuru} : $ \nu_0=\mu_1-C_0/\mu_0= x+1/x-1/x=x$
with $\mu_1=x+1/x$ that follows from \eqref{f1recuru} for $m=0$.
Here we used that $C_0=1$. 

The quantities $f_{1, \, n}= \mu_n,\, f_{2, \, n}=\nu_n,\, f_3^{(n)}= \mu_{n-1}$
 are solutions of the symmetric Painlev\'e 
equations \eqref{symp4} with the parameters
$\alpha_{i, \, n}, i=1,2,3$ given by
\begin{equation}
\alpha_{1, \, n}=C_n\;\; \alpha_2^{(n)}=\frac12 (3-(-1)^n ),\;\;
\alpha_{3, \, n}=C_{-n}\, .
\label{Cmconstant}
\end{equation} 

The constant $C_n$ that appeared in above formulas is defined as:
\begin{equation}
C_n=\frac14 ( 6n+3+(-1)^n )=\begin{cases} 3k+1 &n=2k\\
3k+2 &n=2k+1 \, \, .
\end{cases}
\label{Cmconst}
\end{equation}
Alternatively it can be defined by  the recurrence relation:
\[C_{n+1}-C_n =\frac32-\frac12 (-1)^n=\begin{cases} 1& n=2k\\
2& n=2k+1 \, ,
\end{cases}
\]
with $C_0=1, C_1=2,C_2=4,C_3=5, C_4=7,{\ldots} $ and it is simply 
the sequence of positive integers not divisible by $3$
: $1,2,\boxed{3},4,5,\boxed{6},7,8,\boxed{9},{\ldots} $. In addition 
\[C_{-n}=\frac14 ( -6n+3+(-1)^n )
=\begin{cases} -3 k +1& n=2k\\
-3k-1& n=2k+1 \, .
\end{cases}
\]
and accordingly 
we confirm that 
$\sum_i^3 \alpha_{i, \, n}=3$ with
\[ 
\alpha_{2, \, n}=C_{n+1}-C_n=3-C_n -C_{-n},\;\; \to \;\; C_{n+1}=3-
C_{-n} \, .
\]
Example: 
We here first calculate  $ f_{1, \, 1}, f_{2, \, 1},f_{3, \, 1}$. For 
$m=1$ we obtain 
\[
f_{1, \,1}=x+\frac{1}{x},\;\; f_{2, \,1} = x-\frac{1}{x},\;\; f_{3,
\,1}=x \, .
\]
Inserting these expressions into symmetric Painlev\'e equations
\eqref{symp4}
we find that these ratios of polynomials are solutions 
for
\[ \alpha_{1,\, 1}=C_1=2, \;\; \alpha_{2,\,1}=2,\;\; 
\alpha_{3,\, 1}=C_{-1}=-1 \, .
\]
Note that we can rewrite $f_{1, \,1}$ as a ratio $Q_1/V_1$ with
$Q_1=x^2+1$ and $V_1=x$ being the lowest order Okamoto polynomials
\cite{okamoto}, which will be
derived below in a more systematic way.

\subsection{The recurrence relations for the Volterra case }
Here we will focus  on the case 
of $k=2$ and the basic recurrence relation \eqref{BCuniOmega}.
This relation is equivalent to Volterra equations. For the Volterra 
lattice with proper initial conditions (that justify ignoring
integration constants in the above discussion) 
$\Omega_n$ satisfies the recurrence relations \eqref{BCuniOmega} and
$\mu_n$  reproduced by :
\begin{equation}
\mu_n  =\frac{\Omega_{n+1} \, \Omega_{n-2}}{\Omega_n  \,\Omega_{n-1}}
=\left( \ln \frac{\Omega_n}{\Omega_{n-1}} \right)^{\prime}
, \;\; n=0,1,2,{\ldots} \, ,
\label{undef2}
\end{equation}
satisfies the Volterra equations \eqref{unVolterraeqs}.
 Recently, in reference
\cite{ASVolterra} (see also \cite{AS} and \cite{muller}) this equation was used 
to find determinant solution to Painlev\'e equation
with initial conditions different from what we are using here. We will
comment on determinant expressions that can be associated to equation \eqref{BCuniOmega}
in section \ref{section:wronskians}.

Next, we decompose $\Omega_n$'s into even/odd components:
\begin{equation}
\Omega_n = \omega_n e^{\frac{n x^2}{2}}=
\begin{cases} Q_{k}e^{kx^2} & n=2k , \;\;\qquad k=0,1,2,{\ldots},\\
V_{k} e^{(k-\frac12) x^2} & n=2k-1 , \; \;\; k=0,1,2,{\ldots}, 
\end{cases} \label{Omegadef1}
\end{equation}
where the polynomials $\omega_n$ :
\begin{equation}
\omega_n =\begin{cases} Q_{k} & n=2k , \;\;\qquad k=0,1,2,{\ldots},\\
V_{k} & n=2k-1 , \; \;\; k=1,2,{\ldots}, 
\end{cases} \label{omegadef1}
\end{equation}
satisfy the recurrence :
\begin{equation}
\omega_{n-1} \omega_{n+2}=\omega_n \omega_{n+1}^{\prime}- 
\omega_n^{\prime} \omega_{n+1}+x \omega_n \omega_{n+1}, \;\; n=0,1,2,{\ldots} 
\label{BCuni1}
\end{equation}
In this way we have introduced monic Okamoto
polynomials  \cite{okamoto}, \cite{NouYama1999} here denoted
as $Q_n$ of order : $x^{n(n+1)}+ {\ldots} $
and  $V_n $ of  order : $x^{n^2}+ {\ldots} $.
They are examples of special polynomials associated with the
Painlev\'e equations \cite{umemura}.

Assuming the three initial conditions
\begin{equation} Q_0=1,\;\; V_0=1, \;\; V_1=x \, ,\label{Q0V0V1}
\end{equation}
we are able to reproduce the initial conditions $\mu_0=x,\mu_{-1}=x$.
To verify this we proceed first by obtaining the corresponding
three lowest values of $\Omega_n$'s  :
\begin{equation}\begin{split}
\Omega_{-1} &= V_0 e^{-\frac12 x^2}= e^{-\frac12 x^2}\, ,\\
\Omega_{0} &= Q_0 =1\, ,\\
\Omega_{1} &= V_1 e^{\frac12 x^2 }= x e^{\frac12 x^2}= (e^{\frac12
x^2})^{\prime}\, ,
\end{split}
\label{LowestOms}
\end{equation}
It follows that
$\mu_0=\left( \ln \frac{\Omega_0}{\Omega_{-1}} \right)^{\prime}=x$.
Inserting this into the left hand side of \eqref{undef2} we 
get $\Omega_{-2}=\exp(-x^2)$. This in turn gives the remaining initial
condition
$\mu_{-1}=\left( \ln \frac{\Omega_{-1}}{\Omega_{-2}} \right)^{\prime}=x$.

Inserting $n=0$ into equation \eqref{BCuniOmega} results in
\begin{equation}
\Omega_{-1} \Omega_2= \Omega_0 \Omega_1^{\prime} -\Omega_0^{\prime} \Omega_1
\;\to\; \Omega_2= Q_1 e^{ x^2} =(1+x^2) e^{x^2} \,\to Q_1=x^2+1 \, , 
\label{Qone}
\end{equation}
that provides the unique expression for $\Omega_2$ or equivalently for $Q_1$.
This recursive procedure can easily be extended to higher $n$.
It follows that the  initial conditions \eqref{Q0V0V1} 
reproduce the correct initial condition of Volterra lattice
and allow us to uniquely obtain 
all higher polynomials via the recurrence relations \eqref{BCuniOmega}.

Using ansatz \eqref{Omegadef1} and initial conditions \eqref{LowestOms}
we will consider two sums of derivatives of $\mu_n$. One with the even final limit
and one with the odd.  First consider two equal expressions for the even
summation limit
\begin{equation}
\sum_{p=1}^{p=2n} \mu_p^{\prime} = \sum_{p=1}^{p=2n} \mu_p(\mu_{p+1}-\mu_{p-1})=
\mu_{2n}\mu_{2n+1} -\mu_1 \mu_0 
=\frac{Q_{n-1}Q_{n+1}}{Q_n^2}- Q_1\, ,
\label{sumaup}
\end{equation}
where we used in \eqref{sumaup} that  $\mu_1=Q_1/x, \mu_0=x$ and
relation \eqref{undef2}.

The above sum  can alternatively be calculated as
\begin{equation}
\sum_{p=1}^{p=2n} \mu_p^{\prime} =  
\left(\frac{\Omega_{2n}^{\prime}}{\Omega_{2n}}\right)^{\prime} \, .
\label{sumaup1}
\end{equation}

Next, we consider the same summation but shifted so it ends at the odd
number:
\begin{equation}
\sum_{p=0}^{p=2n-1} \mu_p^{\prime} =  \sum_{p=0}^{p=2n-1} \mu_p(\mu_{p+1}-\mu_{p-1})=
\mu_{2n}\mu_{2n-1} -\mu_0 \mu_{-1}
=\frac{V_{n-1}V_{n+1}}{V_n^2}- x^2
\label{sumaup2}
\end{equation}
and compare it with 
\begin{equation}\begin{split}
\sum_{p=0}^{p=2n-1} \mu_p^{\prime}&= 
 \sum_{p=0}^{p=2n-1} 
\left(\frac{\omega_{p}^{\prime}}{\omega_{p-1}}\right)^{\prime}
+\sum_{p=0}^{p=2n-1} 1\\
&=\left(\frac{V_{n}^{\prime}}{V_{n}}\right)^{\prime} +2 n \, .
\label{sumaup3}
\end{split}
\end{equation}
Combining the recursion relations 
\eqref{sumaup1} and \eqref{sumaup2} we obtain
\begin{equation}
\frac{\Omega_{n-2}\Omega_{n+2}}{\Omega_n^2}= 
\left(\frac{\Omega_n^{\prime}}{\Omega_n}\right)^{\prime}
+ Q_1
, \quad n=1,2,{\ldots}  \, .
\label{FOUomega}
\end{equation}

More explicitly in terms of polynomials $Q_n,V_n$ the above is equivalent to :
\begin{equation}
\frac{Q_{n-1}Q_{n+1}}{Q_n^2}= \left(\frac{Q_n^{\prime}}{Q_n}\right)^{\prime}
+x^2+2 n+1, \quad n=1,2,{\ldots}  \, .
\label{FOUq}
\end{equation}
and
\begin{equation}
\frac{V_{n+1} V_{n-1}}{V_n^2}=\left(\frac{V_n^{\prime}}{V_n}\right)^{\prime}
+x^2+2 n, \quad n=1,2,{\ldots} 
\label{FOUv}
\end{equation}
The recursion relation \eqref{FOUv} only needs two data points to
generate all the remaining higher  polynomials. We choose them as
initial conditions $V_0=1,V_1=x$ and
insert them into the recurrence relation for $n=1$ to produce
$V_{n+1}=V_2= x^4+2 x^2-1 $ and then recursively all the higher polynomials.
For the recursion relation \eqref{FOUq} the similar recursion procedure
involves required data points $Q_0=1$ and $Q_1=x^2+1$ derived via recurrence  relation in \eqref{Qone}
from the initial conditions.

The above recurrence relations \eqref{FOUq} and \eqref{FOUv} 
appeared in \cite{okamoto} and in \cite{FOU},
where they were derived from Hirota bilinear relations of
$\tau$--functions. 
Here we were able to show that they are inherently 
part of the Volterra lattice structure deriving
them  solely from Volterra
relations.

Next we define a new structure:\begin{equation}
\Theta_n^{(1)} =
\omega_{n}^{(1)} e^{n x^2/2}=
\begin{cases} S_{k+1}e^{-kx^2/2} & n=2k , \;\;\qquad k=0,1,2,{\ldots},\\
P_{k} e^{-(k-\frac12 ) x^2/2} & n=2k-1 , \; \;\; k=1,2,{\ldots}, 
\end{cases} \label{Xidef1}
\end{equation}
that satisfy the recurrence relations:
\begin{equation}
\Theta_k \Theta_{k-2}^{\prime} -\Theta_k^{\prime} \Theta_{k-2}=
\Theta_{k-1}\Theta_{k-1}^{(1)}\, , 
\label{SPuni}
\end{equation}
In terms of these quantities we can rewrite $\nu_n$ as  
\begin{equation}
 \nu_n= \frac{\Theta_{n-1} \Theta_{n-1}^{(1)}}{\Theta_n
\Theta_{n-2}}=
(\ln \frac{\Theta_{n-2}}{\Theta_{n}})^{\prime}
\label{mundef}
\end{equation}
that in terms of even/odd components reads
\begin{align}
\nu_{2k-1}&= \frac{Q_{k-1} S_k}{V_k V_{k-1}}= (\ln
\frac{V_{k-1}}{V_k})^{\prime}+x \, , \label{upsilon2km1}\\
\nu_{2k}&= \frac{V_k P_k}{Q_k Q_{k-1}}= (\ln
\frac{Q_{k-1}}{Q_k})^{\prime}+x\, . \label{upsilon2k}
\end{align}

\subsubsection{Connection to the $T_1^{\frac12}$ orbit }
The Okamoto polynomials $Q_n, V_n$ defined in relation \eqref{Omegadef1}
and the generalized Okamoto polynomials 
$P_n,S_n$ defined   in relation \eqref{Xidef1} enter 
the $T_1^{\frac12}$ orbit expressions  for $\mu_n$ and $\nu_n$:
\begin{align}
f_{1,\, 2k}&=\frac{Q_{k-1}V_{k+1}}{Q_kV_k}, \quad
f_{2,\, 2k}=\frac{V_{k}P_{k}}{Q_kQ_{k-1}}, \;\; k=1,2,3{\ldots}, 
\label{fieven}\\
f_{1, \, 2k-1}&=\frac{Q_{k}V_{k-1}}{Q_{k-1}V_{k}}, \quad
f_{2, \,2k-1}=\frac{Q_{k-1}S_{k}}{V_kV_{k-1}}, \;\; k=0,1,2,3{\ldots},
\label{fiodd}
\end{align}
with $f_{3,\, n}=f_{1,\, n-1}$
and  $f_{i,\, n}=T_1^{(\frac{n}{2})}
(f_{1, \, 0},f_{2,\, 0},f_{3,\, 0})$ with 
$(f_{1, \, 0},f_{2, \, 0},f_{3,\, 0})=(x,x,x)$.

If instead of $T_1^{(\frac12)}$ we decided to use
$T_2^{(\frac12)}$, the expressions in \eqref{fieven},
\eqref{fiodd} will be shifted $i \to i+1$ and we would have
the $T_2^{\frac12}$ orbit expressions (with $m=0$) 
\begin{alignat}{2}
f_{2,\, 2k}&=\frac{Q_{k-1}V_{k+1}}{Q_kV_k}, &\qquad\quad
f_{3, \, 2k}&=\frac{V_{k}P_{k}}{Q_kQ_{k-1}}, \;\; k=1,2,3{\ldots}, 
\label{T2fieven}\\
f_{2,\, 2k-1}&=\frac{Q_{k}V_{k-1}}{Q_{k-1}V_{k}}, &
f_{3,\,2k-1}&=\frac{Q_{k-1}S_{k}}{V_kV_{k-1}}, \;\; k=0,1,2,3{\ldots},
\label{T2fiodd}
\end{alignat}
and $f_{1,\, n}=f_{2,\, n-1}$ for
 $f_{i,\, n}=T_2^{(\frac{n}{2})}
(x,x,x)$. Later we will denote $Q_n=V_n^{(-1)}, 
V_n=V_n^{(0)}, P_n=V_n^{(1)}, S_n=V_n^{(2)}$.   

Expressions \eqref{T2fieven}-\eqref{T2fiodd} can be unified as
\begin{equation}
f_{2, \, n}= \frac{\omega_{n+1} \, \omega_{n-2}}{\omega_n  \,\omega_{n-1}},
\qquad\quad
f_{3,\, n}=  \frac{\omega_{n-1}  \, \omega_{n-1}^{(1)}}{\omega_n  \, \omega_{n-2}}
, \;\; n=1,2,3,{\ldots}, 
\label{T2fn}
\end{equation}
where $\omega_n$ and $\omega_{n}^{(1)}$ are defined in 
relations \eqref{omegadef1} and \eqref{Xidef1}.

Further the expressions \eqref{T2fn} can easily  be  generalized to
\begin{equation}
f_{2,\, n}= \frac{\Omega_{n+1} \, \Omega_{n-2}}{\Omega_n  \,\Omega_{n-1}},
\qquad\quad
f_{3, \, n}=  \frac{\Omega_{n-1}  \, \Omega_{n-1}^{(1)}}{\Omega_n  \, \Omega_{n-2}}
, \;\; n=1,2,3,{\ldots}, 
\label{T2fnO}
\end{equation}
with $\Omega_n$ that satisfies the identity \eqref{BCuniOmega}.
Recall that $f_{2,\, n}=\mu_n$ and $f_{3, \,n}=\nu_n$ of the Volterra lattice.

\section{Double  Volterra lattice, generalized Okamoto
polynomials and solutions of the symmetric Painlev\'e IV equations}
\label{section:double}
Here we consider the two-dimensional Volterra lattice structure generated as the
mixed orbit of two different half translations:
\begin{equation}
T_2^{(\frac{n}{2})} T_1^{(\frac{m}{2})} (x,x,x), \quad n\ge 0, \;
m\ge 0
\label{mixedorbit}
\end{equation}

After acting $m$ times with $T_1^{(\frac{1}{2})}$ along the
one-dimensional Volterra generated by $T_1$ we define boundary
conditions for the new Volterra lattice generated by $T_2^{(n/2)}$
stacked on top of the old Volterra lattice generated by $T_1^{(m/2)}$
\begin{align}
f_{2, 0}^{(m)}&=T_1^{(\frac{m}{2})} (x,x,x)\vert_2
\label{b0m}\\
f_{1,0}^{(m)}&=T_1^{(\frac{m}{2})} (x,x,x)\vert_1 \, ,
\label{a0m}
\end{align}
where we projected on $f_2$ and $f_1$ directions, respectively. 
as shown earlier their
Painlev\'e IV parameters are $\alpha_i=\frac12 (1 +3 m,4,1-3m)$ for
$m$ odd and $\alpha_i=\frac12 (2 +3 m,2,2-3m)$ for
$m$ even, respectively.

It is convenient to represent a double Volterra lattice  
by extending
the framework of one dimensional Volterra lattice of eqs. \eqref{f1recuru}-\eqref{f2recuru}
to allow for an additional index. 

We set a new direction in a double  Volterra lattice after the first 
$m$-step in the single Volterra lattice defined by recurrence relations
\eqref{f1recuru}-\eqref{f2recuru} with initial conditions
\eqref{inicmunu}. These $m$ steps yield the final values of $x$
Volterra lattice
\[\mu_m=f_{1, \, n=m}, \;\; \nu_m=f_{2, \, n=m}
, \;\; \mu_{m-1}=f_{3, \, n=m} \, .
\]
The value of $\mu_m$ becomes an initial condition for the double Volterra 
in a new direction with recurrence relations:
\begin{align}
\mu_{n+1}^{(m)}&=\frac12 \left( \frac{(\mu_n^{(m)})'}{ \mu_n^{m}}+3 x -
 \mu_n^{(m)}+ \frac{{\bar  C}_n^{(m)}}{  \mu_n^{m}} \right), \;\; n=0,1,2,{\ldots} 
\label{2D1recuru}\\
\nu_{n}^{(m)} &=  \mu_{n+1}^{(m)} - \frac{{\bar C}_{n}^{(m)}}{ \mu_n^{(m)}} 
, \;\; n=0,1,2,{\ldots}
 \label{2D2recuru}\\
&=  \mu_{n-2}^{(m)}- \frac{{\bar C}_{n-1}^{(m)}}{ \mu_{n-1}^{(m)}} , 
\;\; n=1,2,{\ldots}
\label{2D2arecuru}
\end{align}
with the initial conditions :
\begin{equation}\begin{split}
 \mu_{0}^{(m)}&= \nu_m , \qquad m=1,2,3, {\ldots}, \\
 \nu_{0}^{(m)}&=\mu_{m-1} \, ,
\end{split}
\label{2Dinicond}
\end{equation}
that ``glue'' the one-dimensional 
Volterra structure to
the new  direction of the double Volterra lattice.
On the right  hand sides of relations \eqref{2Dinicond}
 are the final $x$ Volterra objects and on the
left hand side are the first elements of the double Volterra lattice.

Connection with $f_i$'s notation is
\[
\mu_n^{(m)}= f_{2,n}^{(m)}, \quad  
\nu_n^{(m)}=  f_{3,n}^{(m)}, \;\; f_{1,n}^{(m)}=\mu_{n-1}^{(m)}, \quad n=1,2,{\ldots} 
\]
We define the constants ${\bar C}_{n}^{(m)}$ as
\begin{equation}
{\bar C}_n^{(m)}=\left\{ \begin{matrix} 
3k+1, & n=2k, &m=2 l, \\
3k+2 , &n=2k, & m=2l+1\\
3(k-l) +2 , &n=2k+1, & m=2l\\
3(k-l) +1 , &n=2k+1, & m=2l+1 \, .
\end{matrix}
\right. 
\label{barCnmdef}
\end{equation}
In terms of sequence $C_n$ defined in \eqref{Cmconst} it holds
\[ {\bar C}_{2k}^{(2l)}= C_{2k}, \;\;\; 
{\bar C}_{2k}^{(2l+1)}= C_{2k+1}, \;\;\; 
{\bar C}_{2k+1}^{(2l)}= C_{2(k-l)+1}, \;\;\; 
{\bar C}_{2k+1}^{(2l+1)}= C_{2(k-l)} \, .
\]

It follows that in terms of these constants
the parameters of the associated symmetric Painlev\'e IV equations are:
\begin{equation}
\alpha_{1,\, n}^{(m)}=- {\bar C}_{n-1}^{(m)},\qquad 
\alpha_{2,\, n}^{(m)}= {\bar C}_{n}^{(m)} \, ,
\label{alphaCC}
\end{equation}
for all $n,m$.
Thus $\alpha_{3,\, n}^{(m)}=3+{\bar C}_{n-1}^{(m)}-{\bar C}_{n}^{(m)}$.

\[
\alpha_{3,\, n}^{(m)}=3+{\bar C}_{n-1}^{(m)}-{\bar C}_{n}^{(m)}
=\begin{cases} -C_{-(m+1)} &  n-{\rm odd}\\
C_{-m}  &  n-{\rm even} \, .
\end{cases}
\]
As expected for $m=0$:
\begin{equation}
{\bar C}_{n}^{(0)} = C_n\, .
\label{bCnm=0}
\end{equation}
\subsection{Solutions for symmetric
Painlev\'e IV equations obtained by double Volterra lattice}
General solutions are given in terms of  generalized Okamoto
polynomials of the following orders:
\begin{align}
V_n^{(2k)}&=x^{n^2+k(k+1)-nk}+ \ldots , \qquad  n,k=0,1,2,{\ldots}
\label{vn2korder}\\
V_n^{(2k-1)}&=x^{n(n+1)+k^2-nk}+ \ldots   , \qquad  n,k=0,1,2,{\ldots} 
\label{vn2km1order}
\end{align}
The generalized Okamoto polynomials are monic and unique (up to
dihedral symmetry with $x^2 \to  -x^2$ ambiguity) with respect
to their leading terms, meaning that if the leading terms agree then
 whole polynomials agree (due to the recurrence relations, up to $x^2 \to  -x^2$ ambiguity).

For $k=0$ the relations \eqref{vn2korder}, \eqref{vn2km1order}  simplify 
to expressions for      
the  Okamoto polynomials of the single Volterra lattice:
\begin{align}
V_n^{(2k)}\vert_{k=0}&=V_n^{(0)}= x^{n^2}+ \ldots =V_n  , \qquad  n=0,1,2,{\ldots}
\label{vn2k0order}\\
V_n^{(2k-1)}\vert_{k=0}&=V_n^{(-1)}= x^{n(n+1)}+ \ldots  = Q_n , \qquad  n=0,1,2,{\ldots} 
\label{vn2km10order}
\end{align}

For a general case with $m=0,1,2,{\ldots} $ the symmetric  
Painlev\'e IV solutions in terms of $V_n^{(m)}$
polynomials we have obtained within the double Volterra lattice are
given by:
\begin{alignat}{2}
f_{2,\,2n}^{(m)}&=\frac{V_{n-1}^{(m-2)}V_{n+1}^{(m)}}{V_n^{(m-2)}V_n^{(m)}}, &\qquad\quad
f_{3,\,2n}^{(m)}&=\frac{V_{n}^{(m)}V_{n-1}^{(m-4)}}{V_n^{(m-2)}V_{n-1}^{(m-2)}},
\;\; n=0, 1,2,3{\ldots}, 
\label{T2fievenV}\\
f_{2,\,2n-1}^{(m)}&=\frac{V_{n}^{(m-2)}V_{n-1}^{(m)}}{V_{n-1}^{(m-2)} V_{n}^{(m)}}, &
f_{3,\,2n-1}^{(m)}&=\frac{V_{n-1}^{(m-2)}V_{n}^{(m+2)}}{V_n^{(m)}
V_{n-1}^{(m)}}, \;\; n=0, 1,2,3{\ldots},
\label{T2fioddV}
\end{alignat}
with $f_{1,\, n}^{(m)}=f_{2,\,n-1}^{(m)}$ and $3x =\sum_{i=1}^3
f_{i,\, n}^{(m)}$.
These expressions satisfy the symmetric Painlev\'e IV equations with
the parameters \eqref{alphaCC} and are one of the main results of this paper.

With identifications   
\eqref{vn2km1order},\eqref{vn2k0order}, \eqref{vn2km10order}
(see also \eqref{vn2km2order} below) 
we find that $P_n=V_{n}^{(1)}$, $S_n=V_{n}^{(2)}$.
Plugging $m=0$ into expressions
\eqref{T2fievenV}-\eqref{T2fioddV} we recover expressions
\eqref{T2fieven}-\eqref{T2fiodd} obtained from a single 
Volterra structure.

Because of the presence of terms $V_n^{(m-4)}, V_n^{(m-2)}$ on the
right hand sides of equations \eqref{T2fievenV} and \eqref{T2fioddV}
we need to further extend expressions \eqref{vn2korder}-
\eqref{vn2km1order} to $m=-4,-3, -2$. We find by analyzing the underlying
orders of polynomials:
\begin{align}
V_n^{(2k)}\vert_{k=-1}&=V_n^{(-2)}= x^{n^2+n}+ \ldots =Q_n  , \qquad  n=0,1,2,{\ldots}
\label{vn2km1ordera}\\
V_n^{(2k-1)}\vert_{k=-1}&=V_n^{(-3)}= x^{(n+1)^2}+ \ldots  = V_{n+1} , \qquad  n=0,1,2,{\ldots} 
\label{vn2km10ordera}\\
V_n^{(2k)}\vert_{k=-2}&=V_n^{(-4)}= x^{(n+1)^2+1}+ \ldots =
V_{n+1}^{(1)}=P_{n+1}  , \qquad  n=0,1,2,{\ldots}
\label{vn2km2order}
\end{align}

For $m=1$ corresponding to translation
$T_2^{(\frac{n}{2})} T_1^{(\frac{1}{2})} (x,x,x)$ we get from  expressions
\eqref{T2fievenV}-\eqref{T2fioddV}
for $n=2k$ and $n=2k-1$:
\begin{alignat}{2}
f_{2,\, 2k}^{(1)} &=\frac{Q_{k-1}P_{k+1}}{Q_k P_{k}}, &\qquad \quad
f_{3,\, 2k}^{(1)}&=\frac{V_{k}P_{k}}{Q_k Q_{k-1}}
\;\;
k=1,2,3{\ldots},\label{2dim1feven} \\
f_{2,\, 2k-1}^{(1)}&=\frac{Q_{k}P_{k-1}}{Q_{k-1}P_{k}}, 
&
f_{3,\, 2k-1}^{(1)}&=\frac{Q_{k-1}M_{k}}{P_{k} P_{k-1}}\;\; k=1,2,3{\ldots},
\label{2dim1fodd}
\end{alignat}
since it always holds that $f_{1, \, n}^{(m)}=f_{2, \, n-1}^{(m)}$ it 
is not necessary to list these components of solutions explicitly. 

Here 
$S_n=V_n^{(2)}, P_n=V_n^{(1)}$ as seen before and there is a
new polynomial $M_n= V_n^{(3)}$. We give their expressions in Appendix
\ref{section:appendix}.

For the objects $\mu_n^{(m)}$ of double  Volterra we have:
\begin{align}
f_{2, \, 2n}^{(m)}&=\frac{V_{n-1}^{(m-2)}
V_{n+1}^{(m)}}{V_{n}^{(m-2)}V_n^{(m)}}=
\left( \ln (\frac{V_{n}^{(m-2)}}{V_n^{(m)}})\right)^{\prime}+x, \qquad n=1,2,3{\ldots} 
\label{bnm}\\
f_{2, \, 2n-1}^{(m)}&=\frac{V_n^{(m-2)} V_{n-1}^{(m)}}{V_{n-1}^{(m-2)}V_n^{(m)}}
=\left( \ln (\frac{V_{n}^{(m)}}{V_{n-1}^{(m-2)}})\right)^{\prime}+x,
\label{anm}
\end{align}
which satisfy the Volterra equations \eqref{unVolterraeqs} in $n$ for any 
fixed $m \ge 0$. They fit into the setup of equations \eqref{undef2}.

\subsection{Recurrence relations for the generalized Okamoto polynomials}
The recurrence relations for generalized Okamoto polynomials can be
found analogously to what was done for a single Volterra lattice 
and are given by:
\begin{equation}\begin{split}
\frac{V_{n+1}^{(m)} V_{n-1}^{(m)}}{(V_n^{(m)})^2}
&=\left(\frac{(V_n^{(m)})^{\prime}}{V_n^{(m)}}\right)^{\prime}
+x^2+2 n-\frac{m}{2}, \;\;\; m=2 l, \,\, l=0,2,3, \;\;\quad
n=1,2,{\ldots} \\
V_n^{(m=2l)} &= V_l^{(2n-1)}, \;
V_l^{(2n-1)}\vert_{n=0}=Q_l, \;\; V_l^{(2n-1)}\vert_{n=1}=P_l   \\
\frac{V_{n+1}^{(m)} V_{n-1}^{(m)}}{(V_n^{(m)})^2}
&=\left(\frac{(V_n^{(m)})^{\prime}}{V_n^{(m)}}\right)^{\prime}
+x^2+2 n-\frac{m-1}{2}, \;\;\; m=2 l-1, \,\, l=0,1,2, \;\;\quad
n=1,2,{\ldots} \, ,\\
V_n^{(m=2l-1)} &= V_l^{(2n)}, \; \;\;
V_l^{(2n)}\vert_{n=0}=V_l,  \;\;V_l^{(2n)}\vert_{n=1}=S_l 
\label{genOkarecur}
\end{split}\end{equation}
Define 
\begin{equation}
\Omega_n^{(m)}= V_n^{(m)}\times 
\begin{cases} e^{(n -m/4)x^2} &m =2l , l=0,1,2, {\ldots}  \\
e^{(n -(m-1)/4)x^2}&m =2l-1 , l=1,2 , {\ldots} 
\end{cases}
\label{genomeganm}
\end{equation}
then it satisfies the recurrence relations:
\begin{equation}
\frac{\Omega_{n+1}^{(m)} \Omega_{n-1}^{(m)}}{(\Omega_n^{(m)})^2}
=\left(\frac{(\Omega_n^{(m)})^{\prime}}{\Omega_n^{(m)}}\right)^{\prime}
+x^2 \, .
\label{genOmeganmrecur}
\end{equation}
Inserting  $m=-1$ into the second of recurrence relations \eqref{genOkarecur}
we reproduce the recurrence relation \eqref{FOUq} for $V^{(-1)}_n=Q_n$
while inserting  $m=0$ into the first 
of recurrence relations \eqref{genOkarecur} we reproduce the 
recurrence relation \eqref{FOUv} for $V^{(0)}_n=V_n$.
Thus recurrence relations \eqref{genOkarecur} cover all the cases for
arbitrary $m$ and $n$ and agree with recurrence relations:
\[\begin{split}
\Omega_{n-1}^{(m-2)} \Omega_{n+1}^{(m)}&=
(\Omega_{n}^{(m-2)})^{\prime} \Omega_{n}^{(m)}-\Omega_{n}^{(m-2)}
(\Omega_{n}^{(m)})^{\prime}\\
\Omega_{n}^{(m-2)} \Omega_{n-1}^{(m)}&=
(\Omega_{n}^{(m)})^{\prime} \Omega_{n-1}^{(m-2)}-\Omega_{n}^{(m)}
(\Omega_{n-1}^{(m-2)})^{\prime} \,.
\end{split}
\]

\subsection{Duality relations  for generalized Okamoto polynomials}

Recall relations \eqref{vn2korder}, \eqref{vn2km1order}
for the generalized polynomials $V_{n}^{(2k)}$, $V_n^{(2k-1)}$,
that generalize  expressions for 
orders of regular Okamoto polynomials  such
that $Q_{n} = x^{n(n+1)}+{\ldots} , V_{n} = x^{n^2}+{\ldots}$.

These two orders of relations \eqref{vn2korder} and \eqref{vn2km1order}
will agree with each other for $k$ in \eqref{vn2korder} being set to $n$ as given
in \eqref{vn2km1order}, and $n$  in \eqref{vn2korder}
 being set to $k$ as given
in \eqref{vn2km1order}.
This observation leads to the following { duality} identity
\begin{equation}
{V_n^{(2k)}= V_k^{(2n-1)}},\; \;n,k =0,1,2,3,4,{\ldots} \, ,
\label{duality}
\end{equation}
as  expected from equality of the leading terms and verified for all
Okamoto type of polynomials.


{\bf Corollaries:}
The first corollary is:
\[ V_0^{(m)}=\omega_m, \;\; \;\;m=0,1,2,{\ldots} 
\]
where $\omega_n$ is defined in equation \eqref{omegadef1}.

The second corollary:
\[ 
V_n^{(2n)}=V_n^{(2n-1)}\;\; \;\;n=0,1,2,{\ldots} 
\]
For example: $V_0=Q_0$ for $n=0$, $S_1=P_1$ for $n=1$ and
$V_2^{(4)}=V_2^{(3)}$ for $n=2$ etc., 

{\bf Example:}
Consider $V_n^{(6)} \equiv X_n$ and insert it on the left hand side of
relation \eqref{duality} to obtain:
\[
V_0^{(6)}= V_3^{(-1)}=Q_3, \; V_1^{(6)}= V_3^{(1)}=P_3, \;
V_2^{(6)}= V_3^{(3)}, \;V_3^{(6)}= V_3^{(5)}, \;
\]
thus for $V_{n}^{(2k)},  n \le k$ it  is given by known $V_k^{(m)}$ with
$m < 2k$.


Other  corollaries  to the duality relation \eqref{duality} are
relations
\begin{equation}\begin{split}
f_{2,\,2n }^{(2k)}&=f_{3,\,2k }^{(2n+1)} ,
\qquad f_{2,\,2n-1 }^{(2k)}=f_{2,\,2k-1 }^{(2n-1)} \\
f_{2,\,2n }^{(2k-1)}&= f_{3,\,2k-1 }^{(2n)}, \qquad
f_{2,\,2n-1 }^{(2k-1)}= f_{2,\,2k-1 }^{(2n)}
\label{f23dual}
\end{split}
\end{equation}
that show the rules for  commutativity of $m$ and $n$ directions of the double
Volterra
solutions
(modulo $\pi$ transformation).
Let us show how this duality is realized as degeneracy of symmetric
Painlev\'e IV equations.
Consider
\[\begin{split}
\partial_x f_{2,\,2n }^{(2k-1)}&=f_{2,\,2n }^{(2k-1)} (f_{3,\,2n
}^{(2k-1)}- f_{1,\,2n }^{(2k-1)}) +\alpha_{2,\,2n }^{(2k-1)}\\
&=f_{2,\,2n }^{(2k-1)} (f_{3,\,2n
}^{(2k-1)}- f_{2,\,2n-1 }^{(2k-1)}) +\alpha_{2,\,2n }^{(2k-1)} \, .
\end{split}
\]
Applying duality relation \eqref{duality} 
we get
\[\begin{split}
\partial_x f_{3,\,2k-1 }^{(2n)}&=f_{3,\,2k-1 }^{(2n)} (f_{2,\,2k-2
}^{(2n)}- f_{2,\,2n-1 }^{(2k-1)} +\alpha_{2,\,2n }^{(2k-1)}\\
&=f_{3,\,2k-1 }^{(2n)} (f_{2,\,2k-2}^{(2n)}- f_{2,\,2n-1 }^{(2k-1)} 
+\alpha_{2,\,2n }^{(2k-1)}\\
&=f_{3,\,2k-1 }^{(2n)} (f_{1,\,2k-1}^{(2n)}- f_{2,\,2k-1 }^{(2n)}) 
+\alpha_{2,\,2n }^{(2k-1)}\, ,\\
\end{split}
\]
which for
\[
\alpha_{2,\,2n }^{(2k-1)}=\alpha_{3,\,2k-1 }^{(2n)}
\]
will agree with the symmetric Painlev\'e IV equation.
The last identity follows from
\[\begin{split} \alpha_{2,\,2n }^{(2k-1)}&={\bar C}_{2n}^{(2k-1)}=3
n+2 \\&=
{\alpha}_{3,2k-1}^{(2n)}= 3+{\bar C}_{2k-2}^{(2n)}-{\bar C}_{2k-1}^{(2n)}=
3+ 3(k-1)+1-3((k-1)-n)-2\\&=3 n+2 \, .
\end{split}\]

Other consequences of the duality relation \eqref{duality} 
are the following 
identities :
\begin{align}
V_0^{(m)} &=  \begin{cases} V_k^{(-1)}=Q_{k} & m=2k , \;\;\qquad k=0,1,2,{\ldots},\\
V_k^{(0)}=V_{k} & m=2k-1 , \; \;\; k=1,2,{\ldots}, 
\end{cases} \label{V02l}\\
V_{1}^{(m)} &=  \begin{cases} V_k^{(1)}= P_{k}& m=2k , \;\;\qquad k=0,1,2,{\ldots},\\ 
V_k^{(2)}= S_k & m=2k-1 , \; \;\; k=1,2,{\ldots}, 
\end{cases} \label{V12k}\\
V_{-1}^{(m)} &=  \begin{cases}  V_k^{(-3)}=V_{k+1}  & m=2k , \;\;\qquad k=0,1,2,{\ldots},\\
 V_k^{(-2)}=Q_{k} & m=2k-1 , \;\;\qquad k=1,2,{\ldots} \, ,.
 \end{cases} \label{Vm1modd}
 \end{align}

\subsubsection{Degeneracies of the generalized Okamoto Polynomials}

The duality relation \eqref{duality} established  equality between 
$V_n^{(2k)}$ and $V_n^{(2k-1)}$ based on comparison
of the powers
 \[ d_n^{(2k)}=n^2+k(k+1)-nk,\quad
d_n^{(2k-1)}=n(n+1)+k^2-nk
\] 
of their leading terms that agree with each other under 
$n \leftrightarrow k$
substitutions and led to equality \eqref{duality}.

There is additional degeneracy that occurs in the even and
odd sector separately. Because of duality relation it is sufficient 
to consider only $V_n^{(2k-1)}$ with order $d_n^{(2k-1)}=n(n+1)+k^2-nk$
for which we are looking for 
two pairs of integers
$(n,k)$ and $(n_1,k_1)$ such that they satisfy the identity
 $d_n^{(2k-1)}=n(n+1)+k^2-nk 
= n_1(n_1+1)+k_1^2-n_1 k_1 $.

If we set $k=k_1$ then the degeneracy condition is 
\[n+n_1=k-1 \, .\]
Examples:
\[ (n,k)=(0,2), \;  (n,k)=(1,2)\]
with $V_0^{(3)}$ and $V_1^{(3)}$ that are related through 
$V_0^{(3)} (x^2)=-V_1^{(3)} (-x^2)$.

If we set $n=n_1$ then $ k+k_1=n$
 For $n=2$ the solutions are $k=0$ and $k_1=2$ 
with the corresponding polynomials 
\[x^2+1=Q_2=V_1^{(-1)}=-V_2^{(3)} (-x^2)\,.\]
To summarize we have obtained the following degeneracy relation 
\begin{equation} \label{degeneracysummary}\begin{split}
V_n^{(2k-1)}(x) & = C_{n,k}  V_{k-n-1}^{(2k-1)} ({\rm i} x) , \; {\rm for }
\; n<k\\
V_n^{(2k-1)}(x) &=K_{n,k} V_n^{(2(n-k)-1)} ({\rm i} x) , \; {\rm for } \;
n \ge k\, ,
\end{split}
\end{equation}
where the proportionality constants $C_{n,k}, K_{n,k}$ can be a sign
$\pm 1$ or
an imaginary symbol $\pm {\rm i}$. These relations simply reflect degeneracy
caused by dihedral symmetry that in the formalism depending on variable
$x$ is related to the transition $x \to i x$. See our comments in the
introduction on transition $z \to x$ or reference \cite{folding} that
refers to such degeneracy as folding.

It is interesting  to point out that 
the polynomial $V_n^{(n-1)}$ appears unchanged on both sides of relation
\eqref{degeneracysummary} and it is therefore non-degenerated.

\subsection{From duality relations to recurrence relations}

The corollaries  \eqref{V02l}-\eqref{Vm1modd} of the 
duality  relations
provide the necessary initial conditions (at least two data 
points) to the recurrence relations
\eqref{genOkarecur} 
for the generalized Okamoto polynomials thus making these polynomials
unique.

Applying the duality relation to recurrence relations
\eqref{genOkarecur} 
we obtain recurrence relations for the generalized Okamoto polynomials 
with  two different upper arguments $m$:
 \begin{equation}
\frac{V_n^{(m-2)}V_n^{(m+2)}}{(V_{n}^{(m)})^2}=
\left(\frac{(V_{n}^{(m)})^{\prime}}{V_{n}^{(m)}}\right)^{\prime}+
x^2 + 2 \frac{m-1}{2}-n+2 \,,
\label{Vkrecer}
\end{equation}
for $m=0, 1,2,3,{\ldots} $ even or odd.

Examples: Equation \eqref{Vkrecer} for $m=0$ yields
 \begin{equation}
\frac{V_n^{(-2)}V_n^{(2)}}{(V_{n}^{(0)})^2}=
\frac{Q_n S_n}{V_{n}^2}
=
\left(\frac{V_n^{\prime}}{V_n}\right)^{\prime}+
x^2 -n+1 \,,
\label{m0Vkrecer}
\end{equation}
using that in our notation $Q_n=V_n^{(-2)}, V_n =V_{n}^{(0)}$ and
$S_n= V_n^{(2)}$

For $m=1,2$ 
we find from the above recurrence relations 
\eqref{Vkrecer}:
\begin{equation}
\begin{split}
\frac{Q_{n} M_{n}}{P_n^2}&=\left(\frac{P_n^{\prime}}{P_n}\right)^{\prime}+
x^2+2 \frac{m-1}{2}-n+2, \;\;m=1, \;\; n=1,2,..\\
\frac{V_{n} N_{n}}{S_n^2}&=\left(\frac{S_n^{\prime}}{S_n}\right)^{\prime}+
x^2+2 \frac{m}{2}-n+1, \;\;m=2, \;\; n=0,1,2,.
\end{split}
\label{mTodarecur}
\end{equation}
with  $M_n= V_n^{(3)}, N_n= V_n^{(4)}$ in our notation.

The recurrence relations in \eqref{genOkarecur}, \eqref{Vkrecer}
can be combined to generate an algebraic relation 
\[\begin{split}
V_n^{(m-2)}V_n^{(m+2)}&= V_{n+1}^{(m)} V_{n-1}^{(m)}+ (V_n^{(m)})^2
(-3(n- \frac{m}{2}) +1), \;\;\; m=2l\\
V_n^{(m-2)}V_n^{(m+2)}&= V_{n+1}^{(m)} V_{n-1}^{(m)}+ (V_n^{(m)})^2
(-3(n- \frac{m}{2}) +\frac12 ), \;\;\; m=2l-1
\end{split}
\]
For $m=0$, 
and for $m=-1$, recalling that $V_n^{(-3)}=V_{n+1}, V_n^{(-1)}=Q_n$ and 
$V_n^{(1)}=P_n$, 
we get
\begin{equation} \begin{split}
Q_n S_n&= V_{n-1}V_{n+1} + V_n^2 (-3n +1)=V_{n-1}V_{n+1} + V_n^2
C_{-2n}\\
V_{n+1} P_n&= Q_{n-1}Q_{n+1} + Q_n^2 (-3n -1)=Q_{n-1}Q_{n+1} + Q_n^2
C_{-(2n+1)}
\end{split}
\label{QSVP}
\end{equation}
for $n=1,2,3,{\ldots} $.
Thus we can obtain all values of $S_n, P_n, n>0$ from the above
algebraic relations.

It also appears that the orders 
of the following products agree:
\begin{equation} \begin{split}
V^{(2k-2)}_{n} V^{(2k+2)}_{n} & \sim V^{(2k-2)}_{n-1}
V^{(2k+2)}_{n+1}\, , \\ 
V^{(2k-1)}_{n} V^{(2k+3)}_{n} & \sim V^{(2k-1)}_{n-1}
V^{(2k+3)}_{n+1} \, .
\label{doubleduality}
\end{split}
\end{equation}
These relations lead to existence of additional relations of the type:
\begin{equation}
\frac{V^{(m-2)}_{n} V^{(m+2)}_{n}
}{(V_n^{(m)})^2}-\frac{V^{(m-2)}_{n-1}V^{(m+2)}_{n+1}}{(V_n^{(m)})^2}=
C_{m+1}, \;\; m=-2, -1, 0,1,2,{\ldots} 
\label{Cmp1again}
\end{equation}
with the constants $C_n$ given in definition \eqref{Cmconst}.

Let us study these relations in a few basic examples to uncover their 
consequences.

For  $m=-1$ we recall that $V_n^{(-3)} = V_{n+1}, V_n^{(1)} =
V_n^{(3)}=M_n$, $V_n^{(-1)} = Q_n$ and obtain from relation
\eqref{Cmp1again}:
\[
\frac{V_{n+1} P_{n}}{Q_n^2}- \frac{V_n P_{n+1}}{Q_n^2}  =C_{0} = 1 \, .
\]


The solution to the above identities for $P_{n+1}$ is 
\begin{align}
P_{n+1}&= V_{n+1} \left(V_1-\sum_{k=0}^{n}
\frac{Q_k^2}{V_kV_{k+1}}\right)
=  V_{n+1} \left(\frac{P_1}{V_1} -\sum_{k=1}^{n}
\frac{Q_k^2}{V_kV_{k+1}}\right) \nonumber \\
&=  
 V_{n+1} \left(\frac{P_i}{V_i} -\sum_{k=i}^{n}
\frac{Q_k^2}{V_kV_{k+1}}\right), 
\label{Pnid}
\end{align}
for $ i=0,1,2, {\ldots} ,n+1$

As a further example consider \eqref{Cmp1again}   with  $m=0$ with $V_n^{(-2)}=Q_n$,
$V_n^{(2)}=S_n$, $V_n^{(0)}=V_n$:
\[
   \frac{Q_m S_m}{V_m^2}- \frac{Q_{m-1}S_{m+1}}{V_m^2}= C_1=2 ,
\]
with solutions for $S_m$:
\begin{align}
S_{n+1}&= 
Q_{n} \left(\frac{V_2}{Q_1}-2\sum_{k=2}^{n} \frac{V_k^2}{Q_kQ_{k-1}}
-4 \frac{V_1^2}{Q_1}
\right)
,\;\; n=0,1,2,{\ldots} \label{Snid}\\
&= 
Q_{n} \left(\frac{S_1}{Q_0}-2\sum_{k=1}^{n} \frac{V_k^2}{Q_kQ_{k-1}}
\right)=Q_{n} \left(\frac{S_i}{Q_{i-1}}-2\sum_{k=i}^{n} \frac{V_k^2}{Q_kQ_{k-1}}
\right)
,\;\; n=0,1,2,{\ldots}
\end{align}
where we used that 
$S_2=V_2-4 V_1^2$.

The general solution of relation \eqref{Cmp1again}
\begin{equation}
\frac{V^{(m+2)}_{n+1}}{V^{(m-2)}_{n}}-  \frac{V^{(m+2)}_{1}}{V_0^{(m-2)}}
=-C_{m+1} \sum_{k=1}^{n} \frac{(V_k^{(m)})^2}{V_k^{(m-2)}
V_{k-1}^{(m-2)}}
\, , 
 \;\; m=-2, -1, 0,1,2,{\ldots} 
\label{Cmp1again2}
\end{equation}
provides an algebraic expression for $V^{(m+2)}_{n+1}$ 
in terms of $V^{(p)}_{n}$ with $ p \le m$. Thus relation \eqref{Cmp1again2}
serves as an algebraic recursion relation in upper indices of
$V_n^{(m)}$ polynomial.

Note, that the ratio $ V^{(m+2)}_{1}/V_0^{(m-2)}$ on the right hand side 
of relation  \eqref{Cmp1again2} can be given in terms of
polynomials $Q_n,V_n,P_n,S_n$ of the one-dimensional Volterra chain, using the duality
relation \eqref{duality} for $m$ even and  $m$ odd:
\[\begin{split}
m&=2 l: \; \frac{V^{2(l+1)}_1}{V_0^{2(l-1)}}
= \frac{V_{l+1}^{(1)}}{V_{l-1}^{(-1)}}=\frac{P_{l+1}}{Q_{l-1}}, \\
m&=2 l-1: \; \frac{V^{(2l+1)}_1}{V_0^{2(l-3)}}
= \frac{V_{l+1}^{(2)}}{V_{l-1}^{(0)}}=\frac{S_{l+1}}{V_{l-1}}
\end{split}
\]
{\bf Example:} Insert $m=1$ and $n=1$ into relation \eqref{Cmp1again2}
with the result:
\[
\frac{V^{(3)}_{2}}{V^{(-1)}_{1}}-  \frac{S_{2}}{V_0}
=-C_{2}  \frac{(V_1^{(1)})^2}{V_1^{(-1)}\, .
V_{0}^{(-1)}}
\]
After multiplying both sides with $V^{(-1)}_{1}=Q_1$ and using $C_2=4$
we get for 
$V^{(3)}_{2}=M_2$ :
\[
M_2=Q_1 \left( \frac{S_{2}}{V_0}-4  \frac{P_1^2}{Q_1 Q_{0}}
\right)= Q_1 S_2 -4 P_1^2 \, ,
\]
that reproduces expression for $M_2$ given in relations \eqref{Mpolynomials}.
 Inserting $m=2$ and $n=1$ and $C_3=5$ into relation \eqref{Cmp1again2}
gives  the result:
\[ V_2^{(4)}=V_1 P_2 -5 S_1^2= V^{(3)}_{2}=M_2 \, ,
\]
in agreement with the special case of duality \eqref{duality}:
$V_n^{(2n)}= V_n^{(2n-1)}$ for $n=2$.

\section{Negative Volterra lattice}
\label{section:negative}

Define 
\begin{align}
\mu_{-2n}&=  
-\frac{Q_{-n} V_{{-n}+1}}{Q_{{-n}+1}V_{-n}}= 
\frac{Q_{{-n}+1}^{\prime}}{Q_{{-n}+1}}-\frac{V_{-n}^{\prime}}{V_{-n}} 
+x,\;\; n=1,2,{\ldots} \label{anqnvn}\\
\mu_{-(2n-1)}&=
-\frac{Q_{{-n}+1} V_{{-n}-1}}{Q_{{-n}}V_{-n}}=\frac{V_{-n}^{\prime}}{V_{-n}}
-\frac{Q_{{-n}}^{\prime}}{Q_{{-n}}} 
+x, \;\; n=1,2,{\ldots} 
\label{bnqnvn}
\end{align}
in terms of
generalized Okamoto polynomials with negative indices defined in 
section \ref{section:appendix}. We find that $\mu_{-n}$ satisfy
the negative Volterra equations:
\begin{equation}
\frac{\mu_{-n}^{\prime}}{\mu_{-n}}= \mu_{-n+1}-\mu_{-n-1}, \;\; n=0,1,2,{\ldots} \\
\label{nVolterraeqs} \end{equation}

We compare the above expressions with the standard positive Volterra lattice object
$\mu_n$  but 
with the initial  conditions $\mu_0=-x , \mu_{-1}=-x$ 
instead of $\mu_0=-x , \mu_{-1}=-x$.
We find that it holds
\[ \mu_{2n-1}=-\mu_{-2n}, \qquad   \mu_{2n}=-\mu_{-2n-1}, \;\; n=1,2,{\ldots} \, .
\]
Therefore if ${\bar \mu}_n$ satisfies the standard Volterra lattice
\eqref{unVolterraeqs} for $n=1,2,{\ldots} $ with initial conditions
${\bar \mu}_0=-x, {\bar \mu}_{-1}=-x$
then $\mu_{-n}$ defined as
\[
{\bar \mu}_n=-\mu_{-n+1}, \;\; n=1,2,{\ldots} 
\]
indeed satisfies the negative Volterra lattice \eqref{nVolterraeqs}.

For the  Volterra lattice with negative indices
we encounter the  following recurrence  relations
for $Q_{-n},V_{-n}, P_{-n}, S_{-n}$ polynomials:
\begin{align}
Q_{-n}\, V_{{-n}+1} &= W \lbrack \, Q_{{-n}+1}, V_{-n} \,\rbrack -
x\, Q_{{-n}+1}\, V_{{-n}},
\;\;\; n=1,2,{\ldots} 
\label{mWrQnVnm1}\\
Q_{{-n}+1}\, V_{{-n}-1} &= W \lbrack \, V_{-n}, Q_{{-n}} \,\rbrack -
x \,Q_{{-n}}\,
V_{{-n}},\;\;\;
n=1,2,{\ldots} \, ,
\label{mWrQnm1Vnp1}\\
V_{-n} {\bar P}_{-n} &=  W \lbrack \,Q_{-n}, Q_{-n+1}\,\rbrack -x Q_{-n} Q_{-n+1},\;\;\;
n=1,2,{\ldots} \, ,
\label{mWrVmnPmn}\\
(-1)^{n+1} Q_{-n} S_{-n} &=  W \lbrack \,V_{-n-1}, V_{-n-1}\,\rbrack 
-x V_{-n} V_{-n+1},\;\;\;
n=0,1,2,{\ldots} \, ,
\label{mWrVmnSmn}
\end{align}
with initial conditions :
\[Q_0=1,\;\; V_0=1, \;\;  V_{-1}=x \,,\]
that give for  $n=1$:
 \[
 \begin{split}
 Q_{-1}\, V_{0} &= W \lbrack \, Q_{0}, V_{-1} \,\rbrack - x\, Q_{0}\,
 V_{-1}\;\to \;
Q_{-1} = -x^2 +1 
\\
Q_{0}\, V_{-2} &= W \lbrack \, Q_{-1}, V_{-1} \,\rbrack - x \,Q_{-1}\,
V_{-1},\;\;\to V_{-2}=x^4-2x^2-1, \; 
\end{split}
\]
with $W \lbrack \, f, g \,\rbrack= f g^{\prime}- f^{\prime} g $\,.
These expressions agree with 
\[
\mu_{-2}=-  
\frac{Q_{-1} V_{0}}{Q_{0}V_{-1}}= \frac{x^2 -1}{x}, \;\;
\mu_{-3}=-\frac{Q_{0} V_{-2}}{Q_{-1}V_{-1}}=\frac{x^4-2x^2-1}{x(x^2 -1
)}\,,
\]
that agree with the negative Volterra lattice equations \eqref{nVolterraeqs}.

The polynomials $Q_{-n}, V_{-n}$ are related to  
the polynomials $Q_{n}, V_{n}$ shown in Appendix \ref{section:appendix}
in the following way :
\begin{equation}\begin{split}
Q_{-n} (x^2)&= (-1)^{[n/2]}  Q_{n} (-x^2), \qquad  n=0, 1,2,3{\ldots} \\
V_{-2 k} (x^2)&= V_{2k} (-x^2) , \qquad k=0,1, 2,3{\ldots} \\
\left( \frac{V_{-(2 k-1)}}{x}\right) (x^2)&=  
\left( \frac{V_{(2 k-1)}}{x}\right)
(-x^2) , \qquad k=1,2,3{\ldots} 
\label{QVnegn}
\end{split}
\end{equation}
with $ [n/2] = 0,0,1,1,2,2,3,3,{\ldots} $ and accordingly the signs
$(-1)^{[n/2]}$ form a periodic sequence $+,+,-,-
,+,+,-,-,+,+, {\ldots} $.

The negative Okamoto polynomials satisfy the recurrence relations 
\begin{equation}
\frac{Q_{-n-1}Q_{-n+1}}{Q_{-n}^2}= \left(\frac{Q_{-n}^{\prime}}{Q_{-n}}
\right)^{\prime}
+x^2+2(-n)-1, \quad n=1,2,{\ldots}  \, .
\label{nFOUq}
\end{equation}
and
 \begin{equation}
\frac{V_{-n-1}V_{-n+1}}{V_{-n}^2}= \left(\frac{V_{-n}^{\prime}}{V_{-n}}
\right)^{\prime}
+x^2+2(-n), \quad n=0,1,2,{\ldots}  \, .
\label{nFOUvq}
\end{equation}

\section{Generalized Volterra lattice and the $k=3$ case of the 
Bogoyavlensky lattice with special
boundary conditions}
\label{section:k3Bogo}

\subsection{Generalized Volterra structure for $A_4^{(1)}$ Painlev\'e 
equations}
\label{subsection:genvolt}

The $A_4^{(1)}$ symmetric Painlev\'e
equations  are {\cite{noumi,noumibk,Adler-polygon,Adlerchain}}:
\begin{equation}
	f_i^{\prime}= f_i (f_{i+1}+f_{i+3}-f_{i+2}-f_{i+4}) +\alpha_i
	\label{a41eqs}
\end{equation}
for $i=1,2,3,4,5$ (mod$5$) and accompanied by the condition
\begin{equation}
\sum_{i=1}^5 f_i =5 x \, .
\label{conditiona4}
\end{equation}
The class of rational solutions is here constructed out of the 
seed solution
\begin{equation}
f_i^{(0)}=x, \; \alpha_i^{(0)}=1
\label{seeda4}
\end{equation}
by action of the ``fractional'' translation $t_1 = \pi^{-1} s_1$ such that
$t_1^4=T_1=\pi s_4 s_3 s_2 s_1$ with $T_1$ being one of the standard
translation operators of the
$A_4^{(1)}$ invariant Painlev\'e equations. This is just one example of
five fractional translations.

The action of $t_1$ is described by 
a set of generalized Volterra transformations :
\begin{equation} \begin{split}
f_1^{(n+1)}&= f_5^{(n)}, \;\; f_3^{(n+1)}=f_2^{(n)}, \;\; 
f_4^{(n+1)}=f_3^{(n)},\\
f_2^{(n+1)}&= -f_3^{(n)}+f_2^{(n)}+f_4^{(n)}+\frac{(f_5^{(n)})^{\prime}}{f_5^{(n)}}, \;\; 
f_5^{(n+1)}=f_1^{(n)}+f_3^{(n)}-f_2^{(n)} - \frac{(f_5^{(n)})^{\prime}}{f_5^{(n)}},
\end{split}
\label{t1recur1}
\end{equation}
which can also be written as 
\begin{equation} \begin{split}
f_1^{(n+1)}&= f_5^{(n)}, \;\; f_3^{(n+1)}=f_2^{(n)}, \;\; 
f_4^{(n+1)}=f_3^{(n)},\\
f_2^{(n+1)}&= f_1^{(n)}+\frac{C_{n}}{f_5^{(n)}}, \;\; 
f_5^{(n+1)}=f_4^{(n)}-\frac{C_n}{f_5^{(n)}},\ n=0,1,2,{\ldots} 
\end{split}  
\label{t1recur}
\end{equation}
with $ C_1=1,C_2=2,C_3=3, C_4=4,C_5=6, C_6=7,C_7=8,C_8=9, C_9=11 {\ldots} 
$ being the periodic sequence of positive integers with $5$ and
multiples
of $5$ omitted
: $1,2,3,4,\boxed{5},6,7,8,9,\boxed{10},11,{\ldots} $. 

The corresponding $A_4^{(1)}$ Painlev\'e equations for solutions
generated by the \eqref{t1recur} recurrence relations are as follows:
\begin{align}
(f_1^{(n)})^{\prime}&= f_1^{(n)} (f_{2}^{(n)}+f_{4}^{(n)}-f_{3}^{(n)}
-f_{5}^{(n)})- C_n, 
\label{a41f1}\\
(f_2^{(n)})^{\prime}&= f_2^{(n)} (f_{3}^{(n)}+f_{5}^{(n)}-f_{4}^{(n)}
-f_{1}^{(n)})+ d_2^{(n)}, 
\label{a41f2}\\
(f_3^{(n)})^{\prime}&= f_3^{(n)} (f_{4}^{(n)}+f_{1}^{(n)}-f_{5}^{(n)}
-f_{2}^{(n)})+  d_3^{(n)},
\label{a41f3}\\
(f_4^{(n)})^{\prime}&= f_4^{(n)} (f_5^{(n)}+f_{2}^{(n)}-f_{1}^{(n)}
-f_{3}^{(n)})+  d_4^{(n)},
\label{a41f4}\\
(f_5^{(n)})^{\prime}&= f_5^{(n)} (f_1^{(n)}+f_{3}^{(n)}-f_{2}^{(n)}
-f_{4}^{(n)}) +C_{n+1}, 
\label{a41f5}
\end{align}
where $d_i^{(n)}, i=2,3,4, n=1,2,3,{\ldots} $ are periodic sequences:
\[\begin{split}
d_2^{(n)}&=2,1,1,1,2,1,1,1,{\ldots} , \\
d_3^{(n)}&=1,2,1,1,1,2,1,1,{\ldots} , \\
d_4^{(n)}&=1,1,2,1,1,1,2,1,{\ldots} ,
\end{split}
\]
One can verify that
\[ -C_n+d_2^{(n)}+d_3^{(n)}+
d_4^{(n)}+C_{n+1}=5
\]
for $n=1,2,3, {\ldots} $. For the special case 
of $n=0$ and the seed solution we have $C_0=d_2^{(0)}
=d_3^{(0)}=d_4^{(0)}=C_1=1$ and these numbers
as expected sum to $5$.The parameters $C_n,  d_i^{(n)}, i=2,3,4$ are Painlev\'e parameters $\alpha_i$  for solutions of  Painlev\'e
equations on the orbit generated by $t_1$
lattice translations.

The explicit solutions that follow  from transformations 
\eqref{t1recur} can be derived recursively for the lower $n$ :
\begin{equation} \begin{split}
f_1^{(1)}&= x, \;\; f_3^{(1)}=x, \;\; 
f_4^{(1)}=x,\\
f_2^{(1)}&=x+\frac{1}{x}, \;\;
f_5^{(1)}=x-\frac{1}{x},
\end{split}
\label{fn1a4sols}
\end{equation}
and at the next level
\[ \begin{split}
f_1^{(2)}&= x-\frac{1}{x}, \;\; f_3^{(2)}=x+\frac{1}{x}, \;\; 
f_4^{(2)}=x,\\
f_2^{(2)}&=\frac{x (x^2+1)}{x^2-1}, \;\;
f_5^{(2)}=\frac{x(x^2-3)}{x^2-1} \,.
\end{split}
\]
Furthermore
\[ \begin{split}
f_1^{(3)}&= \frac{x(x^2-3)}{x^2-1}, \;\; 
f_3^{(3)}=\frac{x (x^2+1)}{x^2-1}
, \;\; 
f_4^{(3)}=x+\frac{1}{x},\\
f_2^{(3)}&=\frac{x (x^2-1)}{x^2-3}, \;\;
f_5^{(3)}=\frac{x^4-6x^2+3}{x(x^2-3)}\, .
\end{split}
\]
For higher $n$ one encounters ratios of higher polynomials in
$f_i^{(n)}$ expressions connected via recurrence relations 
\eqref{t1recur1}.

One can derive from relations \eqref{t1recur1} and \eqref{t1recur}:
\[ f_5^{(n+1)}= \frac12 \left( 
-\frac{(f_5^{(n)})^{\prime}}{f_5^{(n)}}+ 5 x - f_5^{(n)}+ 2 f_2^{(n)}
-\frac{C_n}{f_5^{(n)}}\right), \;\; f_2^{(n+1)}= f_5^{(n-1)}
+\frac{C_n}{f_5^{(n)}}
\]
for the doublet of $f_2^{(n)},  f_5^{(n)}$ 
after eliminating all the other
components $f_i^{(n)}, i=1,3,4$. These relations are of the type of the Volterra orbit relations
given in equations \eqref{f1recuru}-\eqref{f1recuru} but 
with the crucial substitution of $3x$ by $5x$.

This derivation shows that the rational solutions have been obtained by applying Volterra
structure extended to include additional directions of $A_4^{(1)}$. 
It is expected that other fractional translations of the  $A_4^{(1)}$ 
Painlev\'e model will lead to rational solutions employing similar
Volterra mechanism.

\subsection{Solutions to $k=3$ Bogoyavlensky equations}
\label{subsection:k3Bogo}

We will consider the case of $k=3$ for which the Bogoyavlensky
equations \eqref{Bogoeqs} simplify to equations \eqref{Bogok3}.
We start with setting the initial conditions 
to :
\begin{equation}
F_0=x, \;\; F_{-1}=x, \; F_{-2}=x\, .
\end{equation}
to mimic the dihedral symmetric  solution seen above. We insert these conditions back into the Bogoyavlensky equation \eqref{Bogok3}
for  $n=0$:
\[
\frac{F_0^{\prime}}{F_0}=F_1+F_2-F_{-1}-F_{-2} \;\; \to \;\; 
F_1+F_2= 2 x +\frac{1}{x}
\]
For $n=-1$ we get from  equation \eqref{Bogok3}:
\[
\frac{F_{-1}^{\prime}}{F_{-1}}=F_0+F_1-F_{-2}-F_{-3} \;\; \to \;\; 
F_1= \frac{1}{x}+F_{-3}
\]
We now choose 
\[
F_{-3}= x-\frac{1}{x} \, .
\]
With this choice it follows that 
\[
F_1=x\]
It is easy to check from the Bogoyavlensky equations \eqref{Bogok3} that
also
\[
F_{-4}=x\, .
\]
This setup ensures that  the negative and positive expressions for solutions of 
the Bogoyavlensky equations \eqref{Bogok3} will be very symmetric

For the first few positive solutions of  the Bogoyavlensky equations \eqref{Bogok3}
we find for $F_{n>1}$ corresponding to the above initial conditions:
\[\begin{split}
F_2  &= \,{x}^{-1}+x=\frac{x^2+1}{x}, \qquad \qquad F_3 = \,x \\
F_4 &= \,\frac {-1+2\,{x}^{2}+{x}^{4}}{x \left( 1+{x}^{2} \right) }, \qquad 
\qquad \; \; F_5= \,\frac {3+{x}^{4}+2\,{x}^{2}}{x \left( 1+{x}^{2} \right) } ,
\\
F_6 &= \,{\frac {{x}^{6}+3\,{x}^{4}-3\,{x}^{2}+3}{x \left(
-1+2\,{x}^{2}+{x}^{4} \right) }},
\;\qquad 
F_7
=\frac{(x^2+1)(x^8+4x^6+6x^4-12x^2-3)}{x(-1+2x^2+x^4)(3+x^4+2x^2)}\, .
\\
\end{split}
\]
For negative modes we find in addition to previously quoted $F_{-3},
F_{-4}$ other very symmetric solutions:
\[ F_{-5}= \,\frac {-1-2\,{x}^{2}+{x}^{4}}{x \left( -1+{x}^{2} \right)
},
\quad
F_{-6}= \,\frac {3+{x}^{4}-2\,{x}^{2}}{x \left( -1+{x}^{2} \right)
},\;\;
\;\;
F_{-7}= \,{\frac {{x}^{6}-3\,{x}^{4}-3\,{x}^{2}-3}{x \left(
-1-2\,{x}^{2}+{x}^{4} \right) }} \,, 
\]
and so on. Clearly the choice of the initial conditions ensured such a
symmetry.

\subsection{The $\Omega_n$ recurrence relations for $k=3$ Bogoyavlensky
lattice}

Let us recall the recurrence relation \eqref{BCBogo3} for $k=3$
with
\[F_n  =\frac{\Omega_{n+2} \, \Omega_{n-3}}{\Omega_n  \,\Omega_{n-1}}
=\left( \ln \frac{\Omega_n}{\Omega_{n-1}} \right)^{\prime}
\]
Plugging the values of $F_0, {\ldots} ,F_4$ into \eqref{OmegaF}
we obtain
\[ \begin{split}
\Omega_{-1}&=\Omega_0 e^{-x^2/2}, \qquad  \Omega_1=\Omega_0 e^{x^2/2}\\
 \Omega_2&=\Omega_0 x e^{x^2}, \qquad \Omega_3=\Omega_0 x e^{3 x^2/2}\\
\Omega_4&=\Omega_0 (x^2+1) e^{2 x^2}, \qquad 
\Omega_5=\Omega_0 x^3 e^{5 x^2/2}\\
\end{split}
\]
From the above we recover e.g.
\[ 
F_2=\frac{\Omega_4 \Omega_{-1}}{\Omega_2 \Omega_1}=\frac{x^2+1}{x},
\quad\quad
F_3=\frac{\Omega_5 \Omega_{0}}{\Omega_3 \Omega_2}=x \, .
\]
We now employ the recurrence relation \eqref{BCBogo} to obtain higher
terms  
\[\begin{split}
\Omega_3 \Omega_{-2}&=\Omega_{0}^{\prime}\Omega_{-1}-
\Omega_0\Omega_{-1}{^\prime}  \quad \to \quad 
\Omega_{-2} = \exp(-x^2)\\
\Omega_6 \Omega_1&=\Omega_{4}^{\prime}\Omega_3- 
\Omega_3^{\prime} \Omega_{4} \quad \to \quad 
\Omega_6=(-1+2\,{x}^{2}+{x}^{4}) \exp(3 x^2)\\
\Omega_7 \Omega_2&=\Omega_{5}^{\prime}\Omega_4 -
\Omega_4^{\prime} \Omega_{5} \quad \to \quad 
\Omega_7=x (3+2\,{x}^{2}+{x}^{4}) \exp(7 x^2/2)
\end{split}
\]
To effectively describe the results of the above recurrence relations
we introduce a notation
\[
\Omega_n = Z_n \exp (\frac{n}{2} x^2), \quad
Z_n= x^{P_n}+{\ldots} 
\]
with $Z_n$ being a monic polynomial of order $P_n$ that satisfies 
the recurrence :
\[ P_{n+6}-P_n=n+4
\]
That recurrence generates the whole sequence from the initial six
values:
\[
P_2=1, P_3=1, 
P_4=2, P_5=3, P_6=4,P_7=5 \, .
\]
Indeed $P_8=P_2+6=7, P_9=P_3+7=8$ etc.

Accordingly, the examples of $Z_n$ polynomials are 
\[ 
\begin{split}
Z_7&= x(3+{x}^{4}+2\,{x}^{2}), \quad
Z_{8}=x({x}^{6}+3\,{x}^{4}-3\,{x}^{2}+3), \\
Z_{9}&={x}^{8}+4\,{x}^{6}+6\,{x}^{4}-12\,{x}^{2}-3,\;\;
Z_{10}= x^{10}+7\,x^{8}+14\,x^{6}+42\,x^{4}+21\,x^{2}-21\, .
\end{split}
\]

\subsection{Connection to the $A_4^{(1)}$ symmetric Painlev\'e
equations}

We consider equations \eqref{a41eqs} with conditions \eqref{conditiona4} and 
initial conditions \eqref{seeda4}. 
Inserting the condition $\sum_{i=1}^5 f_i =5 x$ into equation
\eqref{a41eqs} we can rewrite it as a relation :
\begin{equation}
f_{i+1}+f_{i+3}= \frac12 \left( \frac{f_i^{\prime}}{f_i} +5 x -f_i+
 \frac{\alpha_i}{f_i} \right) \, .
\label{a41eqsa}
\end{equation}
The solutions of the Bogoyavlensky equation obtained
above seem to fit into
this scheme.
First observe that 
\[ F_1+F_2= \frac1x +2x = \frac12 \left( \frac{F_0^{\prime}}{F_0} +5 x -
F_0+
 \frac{1}{F_0} \right)\, ,
\]
where $F_0=x$.
Next we reproduce 
\[ F_2+F_3= \frac12 \left( \frac{F_1^{\prime}}{F_1} +5 x -
F_1+
 \frac{1}{F_1} \right) \, ,
\]
and 
\[ F_3+F_4= \frac12 \left( \frac{F_2^{\prime}}{F_2} +5 x -
F_2+
 \frac{2}{F_2} \right) \, .
\]
Generally if holds that
\begin{equation}
F_{n+1}+F_{n+2}= \frac12 \left( \frac{F_n^{\prime}}{F_n} +5 x -
F_n+
 \frac{K_n}{F_n} \right)\, ,
\label{Bogoorbit}
\end{equation}
where
\begin{equation}
\begin{split}
K_0&=1, K_1=1, K_2=2, K_3=3, K_4=4, K_5=4, K_6=6, K_7=6, K_8=7,\\
 K_9&=8,
K_{10}=9, K_{11}=9, K_{12}=11, {\ldots} 
\end{split}
\label{C's}
\end{equation}
This sequence is given explicitly by
\begin{equation} \begin{split}
K_{6m}&=5m+1,\quad K_{6m+1}=5m+1,\quad K_{6m+2}=5m+2,\\
K_{6m+3}&=5m+3,\quad K_{6m+4}= 5m+4, \quad K_{6m+5}=5m+4\, .
\end{split}
\label{Painleveparemeters}
\end{equation}

Also it appears that 
\[
5x-F_{n-2}-F_{n-1}-F_n -F_{n+1}-F_{n+2}=-\frac{K_n}{F_n} \, .
\]
One notices that the quintet $F_{-3}, F_{-2}, F_{-1},F_0, F_1$ agrees with the solutions \eqref{fn1a4sols} to $A_4^{(1)}$ Painlev\'e equations
but for 
increasing $n$ the solutions found by the $k=3$ Bogoyavlensky method differ from those found by the generalized Volterra method that 
reproduced correctly solutions of $A_4^{(1)}$ Painlev\'e equations.

It would be interesting to identify a set of equations based on an independent group theoretic construction that would fit with the Bogoyavlensky lattice structure.
Such goal
will  be pursued in a separate publication.

\section{Discussion and summary  of the results}
\label{section:discussion}

We have established a direct correspondence between
Volterra lattice dynamics and rational solutions of
symmetric 
Painlev\'e 
equations generated by fractions of translation operators.
This correspondence yields explicit closed-form expressions for entire
families of rational solutions, introduces generalized Okamoto
polynomials as natural lattice objects, and produces new algebraic
recurrence relations satisfied by these polynomials. Furthermore, we
discussed an extension beyond the classical
Volterra case to the $k=3$ Bogoyavlensky lattice.{\tiny }

Crucial for this study was the treatment
of initial conditions of Volterra and Bogoyavlensky equations. 
The initial conditions of Volterra equations \eqref{unVolterraeqs}
are of great importance 
and   determine, among other things, whether the lattice consists of 
finite or infinitely many sites.
The two well-known conventional conditions that define finite
Volterra lattices
are open-end conditions ($\mu_0 = \mu_N = 0$, for which  the lattice  effectively
consists of $N - 1$ sites), and
 periodic conditions  (all indices are taken (mod $N$), so 
 that $\mu_0 = \mu_N,
\mu_{N +1}= \mu_1$). More recently, the reference \cite{ASVolterra,AS} 
considered
the Volterra chain with an initial condition equal to $0$ in one node and 
$1$ 
in the others and obtained solutions in terms of Bessel functions.

Our  work is based on condition $\mu_0=\mu_{-1}=x$, that leads to
Volterra lattice with infinitely many 
sites. It 
 first appeared in reference  \cite{kametaka}, in  1983,  to produce solutions
of the Painlev\'e IV equation originating from
the seed solution $y(x)=x$. 
We will refer to such class of
lattices as one-dimensional or single Volterra lattice solution
to distinguish from the Volterra lattice generated by two distinct 
half translations that generate the two-dimensional Volterra lattice.

We have shown how the underlying solutions that start from condition 
 $\mu_0=\mu_{-1}=x$,
are equivalent to  solutions 
$f_{i, \, n}, i=1,2,3$ of  symmetric Painlev\'e IV
equations (with parameters $\alpha_{i,\,n}$ depending on only one
integer parameter
$n= 0,1,2,..$). 
This equivalence 
involves additional sequence of functions $\nu_n$ that
satisfy $\nu_n=3x-\mu_n-\mu_{n-1}$ that leads to the initial condition
$\nu_{0}=x$ and a new equation for $\nu_n$: 
$\mu_n^{\prime}/\mu_n=\nu_n-\nu_{n+1}$.

Plugging $\mu_0=\mu_{-1}=x$ into the Volterra lattice equation
yields $\mu_1=x+1/x$ and so on with  every new element $\mu_n$ being
uniquely determined from Volterra lattice equations and the initial
conditions.
Plugging $\nu_0=x$ into equation for $\nu_n$ yields
$\nu_1=x-1/x$ and so on.
This construction gives rise to  monic Okamoto
polynomials  $Q_n$ of order : $x^{n(n+1)}+ {\ldots} $
and  $V_n $ of  order : $x^{n^2}+ {\ldots} $ for $n=0,1,2, {\ldots} $.

The explicit correspondence between the one-dimensional Volterra lattice  and symmetric
Painlev\'e IV solutions is given by 
$f_{i, \, n}= \mu_n$, $f_{i+1, \, n}= \nu_n, f_{i-1, \, n}= \mu_{n-1}$
for any $i$ with values $1,2,3$ (mod $3$).

There also exists a negative  Volterra lattice 
with the boundary conditions $\mu_0=\mu_{-1}=-x$. Such negative
hierarchy is related  to the transformation $x^2 \to -
x^2$ of arguments of the Okamoto polynomials. This transformation is a residual
symmetry of a dihedral symmetry $D_3$ of the symmetric Painlev\'e IV
equations.

The crucial observation for this study is that for any finite
positive integer $m$ the
objects   $\mu_m, \nu_m$
of the $+x$ Volterra hierarchy can be used as an initial
conditions,
$\mu^{(m)}_0=\nu_m,\, \nu^{(m)}_0=\mu_{m-1}$,
for another independent Volterra lattice 
stacked on top of the one-dimensional Volterra lattice.
Such Volterra lattice with  general 
solutions $f_{i, \, n}^{(m)}= \mu_n^{(m)}$, $f_{i+1, \, n}^{(m)}=
\nu_n^{(m)}$ 
of the   symmetric Painlev\'e IV
equations will depend on  two independent integers (reproducing 
the maximum number
of independent parameters that $\alpha_i, \,i=1,2,3$
with the condition $\sum_{i=1}^3 \alpha_i=3$ can depend on). 
We refer to such Volterra
lattice as a double or two-dimensional 
Volterra lattice because of its initial conditions
and the fact that two Volterra lattices are involved in its construction.
This lattice is generated by two independent
square-roots of translation operators associated with two different
directions of the $A^{(1)}_2$ group.

Double Volterra lattice is conveniently described in terms of
generalized Okamoto polynomials $V^{(m)}_n (x),\, m=-1, 0,1,2,{\ldots} ,\,
n=0, 1, 2, {\ldots} $ In terms of these polynomials the boundary
conditions of the double Volterra lattice are given by 
$V_{-1}^{(2n)}=V_{n+1}, V_{-1}^{(2n+1)}=Q_n$ and $V_{0}^{(2n)}=Q_{n},
V_{0}^{(2n+1)}=V_n$ with the standard Okamoto polynomials $Q_n, V_n, \,
n=0,1,2, {\ldots} $. 
The remaining values of generalized Okamoto
polynomials $V^{(m)}_n (x)$ are then obtained via appropriate recurrence
relations.


One of a long term objectives of the project is  to find a general connection between
polygons with dihedral symmetry and polynomials generated 
by such construction. Given a polynomial can you guess what polygon it
came from?
Another long term objective is to describe all the details of
the $k>2$ Bogoyavlensky 
lattices and their generalized polynomial structure.

\appendix
\section{Few concrete expressions for Okamoto Polynomials}
\label{section:appendix}
The standard Okamoto polynomials $V_n^{(-1)}=Q_n, V_n^{(0)}=V_n, 
n \in   \mathbb{Z}$ are of orders
$\vert n \vert (\vert n \vert +1)$ and $n^2$, respectively.
We list their values for several positive and negative $n$: 
\begin{equation} \begin{split}
 Q_{-4} &= x^{20}-30\, x^{18}+355\, x^{16}-2200\, x^{14}+8050\,
 x^{12}-20020\, x^{10}\\&+
 42350\, x^8-107800\, x^6+202125\, x^4-134750\, x^2+67375\\
Q_{-3}&= -x^{12}+14\, x^{10}-65\, x^8+140\, x^6-175\, x^4+350\, x^2-175\\
Q_{-2} &= 5\, x^2-5+x^6-5\, x^4, \;\;\;Q_{-1} = -x^2+1,\\
Q_0&=1, \;\;  Q_1= x^2+1,  \;Q_2=x^6  + 5 x^4  + 5 x^2 + 5,\\
Q_3&=x^{12}+14 x^{10}+65 x^8+140 x^6+175 x^4+350 x^2+175\\
 Q_{4} &= x^{20}+30\, x^{18}+355\, x^{16}+2200\, x^{14}+8050\,
 x^{12}+20020\, x^{10}\\&+
 42350\, x^8+107800\, x^6+202125\, x^4+134750\, x^2+67375\\
\end{split}
\label{QQs}
\end{equation}
and
\begin{equation} \begin{split}
 V_{-4} &= x^{16}-20\, x^{14}+140\, x^{12}-420\, x^{10}+350\, x^8+980\, x^6
 \\&-4900\, x^4+4900\, x^2+1225,\;\; \; \; 
V_{-3} = x(-35+14\, x^4+x^8-8\, x^6)\\
V_{-2}&= -1+x^4-2\, x^2, \;\; V_{-1}=x, \;\;
V_0=1, \;\; V_1=x, \;\;  V_2= x^4+2 x^2-1,\\
V_3&= x(x^8   + 8 x^6 + 14 x^4  - 35), \;\;\;
V_4=x^{16}+20 x^{14}+140 x^{12}+420 x^{10}\\&+350 x^8-980 x^6-4900 x^4-
4900 x^2+1225
\end{split}
\label{VVs}
\end{equation}
For $S_n=V_n^{(2)},P_n=V_n^{(1)}$ we find from the relevant recurrence relations:
\begin{equation}\begin{split}
S_m&:\, S_{-3}= x^{14}-7x^{12}-21x^{10}+175 x^8-245
x^6-245x^4-735x^2+245,\\
&S_{-2}=x^8-14 x^4-7,\, S_{-1}=x^4+2x^2-1,\,S_0=x^2+1,
\,S_1=x^2-1, \\
&S_2=x^4-2x^2-1, \; S_3 = x^8-14 x^4-7\\
P_m&: P_{-3}=x^{10}-5 x^8-10x ^6+50x^4-75x^2 
-25, \; P_{-2}=x (x^4-5),\\ &P_{-1}=x^2+1,\;
P_{0}=x 
, \;P_1=x^2-1,\; P_2=x (x^4-5),\\&
P_3=x^{10}+5 x^8-10x ^6-50x^4-75x^2 
+25\\
&P_4 = x(x^{16}+16x^{14}+60x^{12}-160x^{10}-1650x^8-4400x^6-7700 x^4+9625)
\end{split}
\label{SPvalues}
\end{equation}

The above polynomials with negative indices are obtained as follows
from  polynomials with positive indices 
\[\begin{split}
\left( \frac{P_{-2 k}}{x}\right) (x^2)&=  
\left( \frac{P_{2 k}}{x}\right)
(-x^2) , \qquad k=0, 1,2,3{\ldots} \\
P_{-(2k-1)} (x^2)&=-P_{2k-1} (-x^2)  , \qquad k=1,2,3{\ldots} \\
S_{-n}(x^2)&=- (-1)^{[(n+1)/2]} S_{n+1} (-x^2), \quad n=0,1,2,{\ldots},
\end{split}
\]
where $-(-1)^{[(n+1)/2]}=-,+,+,-,-,+,+,{\ldots} $ for $n=0,1,2,{\ldots} $.

For $M_n= V_n^{(3)}$ we find from relations \eqref{V02l}, \eqref{V12k}
and \eqref{Vm1modd} that $M_0=V_2, M_1= S_2, M_{-1}=Q_2$, which after
plugging into recurrence relations \eqref{recurMpol} allows us to find 
the remaining values of $M$-polynomials
\begin{equation} \begin{split}
M_{-3}&=x^{16} -60x^{12}+550x^8-5500 x^4-1375\\
M_{-2}&=x^{10}+5 x^8-10x^6-50x^4-75x^2+25\\ 
M_{-1} &=5 \,x^2+5+x^6+5\,x^4, \;\; M_0=x^4+2\,x^2-1\\
M_1 &= -1+x^4-2\, x^2,\;\; M_2= 5\, x^2-5+x^6-5\, x^4,\\
M_3 &=x^{10}-5 x^8-10x^6+50x^4-75x^2-25\\
M_4&=x^{16} -60x^{12}+550x^8-5500 x^4-1375
\end{split}
\label{Mpolynomials}
\end{equation}
They satisfy the recurrence relations
\begin{equation}
 \frac{M_{n-1}M_{n+1}}{M_{n}^2}=
 \left(\frac{M_{n}^{\prime}}{M_{n}}\right)^{\prime}
+x^2+2 n-1, \quad n=1,2,{\ldots}  \, .
\label{recurMpol}
\end{equation}
that gives unique answers due to 
initial conditions $M_1=S_2=x^4-2x^2-1$, $M_{-1}=Q_2$.
as follows from relations \eqref{genOkarecur}.

The $M$-polynomials with negative indices are obtained 
through relations $M_{-3} (x^2)=M_4 (-x^2)$, $M_{-n} (x^2)=-M_{n+1}
(-x^2),n=0,1,2$. Since $M_n (x^2)=(x^2)^{n(n-1)/2+2}+\ldots $
for $n$ such that $n(n-1)=4 k, k  \in \mathbb{N}$ e.g. 
$n=4,5,8,9,12,13,{\ldots} $ we will have
$M_n (-x^2)=M_n (x^2)$.

For $N_n= V_n^{(4)}$ we get 
\begin{equation} \begin{split}
N_{-1}&= x\,(x^8+8\,x^6+14\,x^4-35),\;\; N_0=5\, x^2+5+x^6+5\, x^4,\\
N_1&= x \,(x^4-5),\;\;\; N_2= 5\, x^2-5+x^6-5\, x^4,\\
N_3 &=x\,(x^8-8\,x^6+14\,x^4-35)\\
N_4&=x^{14}-7\,x^{12}-21\,x^{10}+175\,x^8-245\,x^6-245\,x^4-735\,x^2+245
\end{split}
\label{Npolynomials}
\end{equation}
that satisfy the recurrence 
\begin{equation}
 \frac{N_{n-1} N_{n+1}}{N_{n}^2}=
 \left(\frac{N_{n}^{\prime}}{N_{n}}\right)^{\prime}
+x^2+2 n-2, \quad n=1,2,{\ldots}  \, .
\label{recurNpol}
\end{equation}
as follows from relations \eqref{genOkarecur}.

\section{Generalized Okamoto polynomials in terms of Wronskians}
\label{section:wronskians}
For completeness we provide Wronskian solutions of 
the recurrence equations \eqref{BCuniOmega}.

Define \cite{AGZ-AIP}:
\[ G_0= \Omega_1=x e^{x^2/2}, \quad F_0=\frac{1}{\Omega_{-1}}=
\int^x \Omega_1= e^{x^2/2} \, ,
\]
  in terms of quantities introduced in relations \eqref{LowestOms}.
Next define 
\[ G_n = \dder[3]{G_{n-1}}{x}=\dder[3n]{G_{0}}{x}, \quad 
F_n=\dder[3]{F_{n-1}}{x}=\dder[3n]{F_{0}}{x},\;\; n=1,2, ..
\]
Then we calculate the following Wronskians 
\begin{align}
V_{n}&= {\rm constant}\times e^{- \frac{n}{2} x^2} W[G_0,G_1,{\ldots}
, G_{n-1}],  \;
n=1,2{\ldots} \nonumber\\
Q_{n}&= {\rm constant}\times e^{- \frac{n+1}{2} x^2}
W[F_0,F_1,{\ldots} , F_n],  \;
n=1,2{\ldots} \nonumber
\end{align}
where constants are chosen so that  the polynomials are monic.

Furthermore, it holds that
\begin{align}
V_{-n}^{(2k+2)}&={\rm constant}\times e^{- \frac{2n+2+k}{2} x^2} 
W[F_0,F_1,{\ldots} ,
F_{n+1+k},G_0,G_1,{\ldots} , G_{n-1} ],  \; \label{Vmn2kp2}\\
n&=1,2{\ldots}, \quad k=-1,0,1,2,{\ldots}  \nonumber\\
V_{-n}^{(2k-1)}&={\rm constant}\times e^{- \frac{2n+2-k}{2} x^2} 
W[F_0,F_1,{\ldots} ,
F_{n-k},G_0,G_1,{\ldots} , G_{n} ],  \;\label{Vmn2km1}\\
n&=1,2{\ldots}, \quad k=0,1,2,{\ldots} \, , \nonumber
\end{align}
from which we can easily obtain expressions for $V_{n}^{(2k+2)},
V_{n}^{(2k-1)}$, As before the constants are chosen to ensure that the
generalized Okamoto polynomials are monic.

There are several formulations of generalized Okamoto polynomials in the
literature starting with \cite{NouYama1999} and including 
\cite{Clarkson-jmp},  \cite{clarksonschrodinger}, 
\cite{islam}, \cite{yangyang},  \cite{JKM}, \cite{kajiwaraohta},
\cite{stokes} and others. These approaches  often derive the polynomials from
Hankel determinants and  Schur functions for the Painlev\'e IV system and its
generalizations 
\cite{kajiwaraohta,JKM},\cite{clarksonfilipuk} but their comparison
often  requires 
relabeling of indices and 
rescaling of coordinates. 
{In reference \cite{clarksonfilipuk} the solutions to
$A_{2n}^{(1)}, n>1$ 
are obtained  from a set of conjectures that generalize the 
expressions in terms of Wronskians  of Schur polynomials, 
given in \eqref{OkamotoPoly1} for the case case of $n=1$ to the higher
$n$. It is important to compare the results of these 
conjectures with the results for $A_{4}^{(1)}$ obtained below from the
fractional translation operators and the generalized Volterra lattices.}

For an  approach towards a qualitative 
classification of the real
solutions of symmetric Painlev\'e IV equations
for suitable parameter values see \cite{twiton2}.

The representation of generalized Okamoto polynomials
that fits well  with notation of our formalism is the 
one given in  \cite{yangyang}.

Let $p_{j}(z)$ be Schur polynomials defined by
\[
\sum_{j=0}^{\infty}  p_{j}(x) \epsilon^j= \exp \left(x \epsilon +
\frac12 \epsilon^2 \right),
\]
with $p_{j}(x)\equiv 0$ for $j<0$.

In terms of Wronskians  of Schur polynomials, the reference \cite{yangyang}
defines
polynomials as
\begin{equation}\label{OkamotoPoly1}
Q_{n_1, \,n_2}(x) = 
\mbox{W}[p_2, p_5, \cdots, p_{3n_1-1}, p_1, p_4, \cdots, p_{3n_2-2}],
\end{equation}

Then the connection between \cite{yangyang} and the generalized
Okamoto polynomials 
obtained here in  Volterra formalism for the odd case is as follows:
\begin{equation}
V_n^{(2k-1)} (x) = c Q_{n,k} (x) \, ,
\label{Yangconnection}
\end{equation}
where $c$ is a constant chosen to ensure that $V_n^{(2k-1)} (x)$ is a monic
polynomial. The order of the polynomial is $d_n^{(2k-1)} = n(n+1) +k^2-nk$. 

All the above generalized Okamoto polynomials are of the odd case as
in equation \eqref{Yangconnection}.
However from the duality relation \eqref{duality} we obtain
\[
c  Q_{k,n} (x)= V_k^{(2n-1)} (x)= V_n^{(2k)} (x)\, .
\]
So even polynomials are obtained just by commuting indices of
$Q_{p\,, r}$.

\vskip .4cm 
\subsection*{Acknowledgements}

JFG thank CNPq and FAPESP for support. YFA thanks FAPESP for financial support
under grant \#2022/13584-0. GVL thanks FAPESP for financial support 
under grant \#2024/16787-4.

\label{lastpage}

\begin{thebibliography}{99}
\bibitem{Adler-polygon}
{V. E. Adler 
\textit{Functional Analysis and Its Applications}  {\bf 27}:2,
141--143, 1993. }
\bibitem{Adlerchain}
{V.E. Adler, \textit{
Physica D: Nonlinear Phenomena} {\bf 73} (4), 335-351, 1994.}
\bibitem{ASVolterra} 
V. E. Adler, A. B. Shabat, \textit{Teoreticheskaya i Matematicheskaya
Fizika}
\textbf{201}:1   37-53, 2019. 
\bibitem{AS}
V.E. Adler, A.B. Shabat,  
\textit{Teoreticheskaya i Matematicheskaya Fizika }\textbf{201}:1
1442-1456, 2019.
V. E. Adler, A. B. Shabat, 
\textit{JETP Letters} \textbf{108}:12  825-828, 2018.
V.E. Adler, \textit{Russ. J. Math. Phys.}, \textbf{31}:1 , 1, 2024.
\bibitem{AAGZ} V.C.C. Alves, H. Aratyn, J. F. Gomes, and A. H. Zimerman,
\textit{J. Phys. A: Math. Theor.} \textbf{53}:44 445202, 2020.
 
\bibitem{AFGZ}
H.~Aratyn, L.~A.~Ferreira, J.~F.~Gomes and A.~H.~Zimerman,
\textit{Phys.\ Lett.}\  B {\bf 316} 85, 1993. 

\bibitem{AGZ-AIP}
H. Aratyn, J. F. Gomes, and A. H. Zimerman, 
\textit{AIP Conference Proceedings} \textbf{1212} 146, 2010.
doi:10.1063/1.3367030,


\bibitem{AGLZ-braids}
H. Aratyn, J. F. Gomes, G. V. Lobo, A. H. Zimerman,
in ``Tribute to Ruben Aldrovandi''  F. Caruso, J. G. Pereira, 
A. Santoro, Editors, pp 121--134, S\~{a}o Paulo, 
Livraria da F\'{\i}sica, 2024

\bibitem{p5degegneracy}
H. Aratyn, J. F. Gomes, G. V. Lobo, A. H. Zimerman,
\textit{Mathematics } \textbf{12}(23) 3701, 2024.


\bibitem{Bogo}
O.I. Bogoyavlensky,
\textit{Russ. Math. Surveys} 46:3  1-64, 1991.


\bibitem{muller}
X.M. Chen, X.B. Hu and F.  M\"uller-Hoissen,
\textit{Nonlinearity} \textbf{31} 4393, 2018.

\bibitem{Clarkson-jmp}
P. A. Clarkson, 
\textit{J. Math. Phys.} {\bf 44}, 5350--5374,  2003.
    \bibitem{clarksonschrodinger}
 P. A. Clarkson, 
\textit{European J. Appl. Math.} \textbf{17}  no.~3, 293--322, 2006.
    \bibitem{clarksonfilipuk}
G.V. Filipuk and P.A. Clarkson, 
\textit{Studies in Applied Mathematics} \textbf{121}: 157-188, 2008.

\bibitem{FOU}
S. Fukutani, K. Okamoto and H. Umemura,
\textit{Nagoya Math. J.},  {\bf 159}  179--200, 2000.


\bibitem{islam}
S.P. Islam, \emph{Special polynomials associated with rational solutions of 
Painlev\'e  equations}, https://doi.org/10.22024/UniKent/01.02.100768

 \bibitem{JKM}
N. Joshi, K. Kajiwara and M. Mazzocco,
\textit{Funkcial. Ekvac} \textbf{49}  451--468,  2006.


\bibitem{kajiwaraohta}
K. Kajiwara and Y. Ohta,
\textit{J. Phys. A } \textbf{31}  no.~10, 2431-24, 1998.


\bibitem{kametaka}
Y. Kametaka,
\textit{Proc. Japan Acad. Ser. A} Math. Sci.\textbf{59} (10): 453-455, 1983.




\bibitem{noumi}  
 M. Noumi and Y. Yamada,  \textit{Communications in Mathematical Physics}
{\bf 199} 281-295, 1998.


\bibitem{NouYama1999}
 M. Noumi and Y. Yamada, 
\textit{Nagoya  Math. J.} \textbf{153} 53 - 86, 1999. 


\bibitem{noumibk}
{M. Noumi,  Painlev\'e Equations through Symmetry, in: Translations of
Mathematical Monographs, vol. 223, (American Mathematical Society
Providence, RI), 2004.}

\bibitem{okamoto}
K. Okamoto,  \textit{Math. Ann.} \textbf{275} 221-255, 1986.


 
\bibitem{stokes}  
P. Roffelsen and A. Stokes, 
\textit{ J. London Math. Soc.}, \textbf{112}: e70329,  2025.  


  \bibitem{twiton2}
 J. Schiff and M. Twiton,
 \textit{Mathematics} \textbf{12} , no.~3, 463, 2024.

\bibitem{folding}
T. Tsuda, K. Okamoto and H. Sakai, 
\textit{Math. Ann.} {\bf 331} 713-738, 2005.

\bibitem{hietarinta}
R. Willox and J. Hietarinta,
\textit{J. Phys.} \textbf{A 36}, 42 , 10615-10633,  2003.



\bibitem{hietarinta1}
{R. Willox and J. Hietarinta,
 On the bilinear forms of Painlev\'e's  4th equation, in  ``Bilinear
Integrable Systems: From Classical to Quantum, 
Continuous to Discrete'' (NATO Science Series
II Mathematics Physics Chemistry vol 201) (Dordrecht: Springer) 
pp 375--90, 2006.}


\bibitem{umemura}
H. Umemura, 
\textit{ Annales de la facult\'e des sciences de Toulouse
 Math\'ematiques, 
 Tome} \textbf{XXIX}, no 5  p. 1063-1089, 2020. 

 \bibitem{yangyang}
B. Yang and J. Yang, 
\textit{ Phys. Lett. A } \textbf{458}  128573, 2023.
\end{thebibliography}
\end{document}